\documentclass{aa}
\usepackage{support-caption}
\usepackage{graphicx}
\usepackage{txfonts}
\usepackage{amssymb,amsfonts,amsmath,bm}
\usepackage{epstopdf,epsfig,subfig,float}
\usepackage{hyperref}
\usepackage{xcolor,color}
\usepackage{natbib}
\usepackage{soul}
\usepackage[english]{babel}
\usepackage{placeins}
\usepackage[normalem]{ulem}

\makeatletter
\renewcommand*\aa@pageof{, page \thepage{} of \pageref*{LastPage}}
\makeatother

\newcommand{\pD}[2]{\frac{\partial #2}{\partial #1}}
\newcommand\bb[1]{\mbox{\boldmath{$#1$}}}
\newcommand\grad{\bb{\nabla}}
\newcommand\bcdot{\,\bb{\cdot}\,}
\newcommand\btimes{\,\bb{\times}\,}

\newcommand\bs[1]{\boldsymbol{#1}}

\definecolor{darkblue}{rgb}{0.0,0.0,0.4}
\hypersetup{colorlinks,breaklinks,
            linkcolor=darkblue,urlcolor=darkblue,
            anchorcolor=darkblue,citecolor=darkblue}
\definecolor{kulblue}{HTML}{116E8A}

\begin{document}

\title{Bridging hybrid- and full-kinetic models\\with Landau-fluid electrons}

\subtitle{I. 2D magnetic reconnection}

\author{F. Finelli\inst{1} \and
        S. S. Cerri\inst{2} \and
        F. Califano\inst{1} \and
        F. Pucci\inst{3,4} \and
        D. Laveder\inst{5} \and
        G. Lapenta\inst{3} \and
        T. Passot\inst{5}}

\institute{Physics Department ``E. Fermi'', University of Pisa, Largo Bruno Pontecorvo 3, 56127 Pisa, Italy\\
        \email{francesco.finelli@phd.unipi.it} \and
        Department of Astrophysical Sciences, Princeton University, 4 Ivy Lane, Princeton, NJ 08544, USA\and
        Centre for Mathematical Plasma Astrophysics, Department of Mathematics, KU Leuven, Celestijnenlaan 200B, 3001 Leuven, Belgium\and
        Istituto per la Scienza e Teconologia dei Plasmi, Consiglio Nazionale delle Ricerche (ISTP-CNR), Via Amendola 122/D, 70126 Bari, Italy\and
        Universit\'e C\^ote d'Azur, Observatoire de la C\^ote d'Azur, CNRS, Laboratoire J.L. Lagrange, Bd. de l'Observatoire, C.S. 34229, 06304 Nice, France}

\date{Received \today; accepted -}

\abstract
{Magnetic reconnection plays a fundamental role in plasma dynamics under many different conditions, from space and astrophysical environments to laboratory devices. High-resolution {\em in situ} measurements from space missions allow naturally occurring reconnection processes to be studied in great detail. Alongside direct measurements, numerical simulations play a key role in the investigation of the fundamental physics underlying magnetic reconnection, also providing a testing ground for current models and theory. The choice of an adequate plasma model to be employed in numerical simulations, while also compromising with computational cost, is crucial for efficiently addressing the problem under study.
}
{
We consider a new plasma model that includes a refined electron response within the "hybrid-kinetic framework" (fully kinetic protons and fluid electrons).
The extent to which this new model can reproduce a full-kinetic description of 2D reconnection, with particular focus on its robustness during the nonlinear stage, is evaluated.}
{We perform 2D simulations of magnetic reconnection with moderate guide field by means of three different plasma models: (i) a hybrid-Vlasov-Maxwell (HVM) model with isotropic, isothermal electrons, (ii) a hybrid-Vlasov-Landau-fluid (HVLF) model where an anisotropic electron fluid is equipped with a Landau-fluid closure, and (iii) a full-kinetic model.
}
{
When compared to the full-kinetic case, the HVLF model effectively reproduces the main features of magnetic reconnection, as well as several aspects of the associated electron microphysics and its feedback onto proton dynamics. This includes the global evolution of magnetic reconnection and the local physics occurring within the so-called electron-diffusion region, as well as the evolution of species' pressure anisotropy. 
In particular, anisotropy-driven instabilities (such as fire-hose, mirror, and cyclotron instabilities) play a relevant role in regulating electrons' anisotropy during the nonlinear stage of magnetic reconnection. As expected, the HVLF model captures all these features, except for the electron-cyclotron instability.
}
%
{}




\maketitle
\section{Introduction}\label{sec:intro}

Weakly collisional, magnetized plasma dynamics are characterized by a wide range of scales, both in space and time. Moreover,  the nonlinearities unavoidably at play in collisionless plasma dynamics can simultaneously couple multiple scales within the system. 
On top of that, plasma processes taking place at kinetic scales are able to provide important feedback on the global (i.e., fluid-scale) evolution of a system, for example in the context of magnetic reconnection (MR). Reconnection is probably the best example of a nonideal, microscale process able to change the global magnetic-field topology of a system by locally breaking the magnetic connections ideally preserved at magnetohydrodynamic (MHD) scales~\citep{BiskampPhR1994}.
The change in magnetic-field connections occurring in relatively short times (to the extent that it is sometimes defined as ``explosive'') in turn triggers the release of a huge amount of energy stored in the previous magnetic configuration~\citep[e.g.,][and references therein]{TreumannBaumjohannFrP2013}.
As for MR, several other processes exist that can lead to the coupling between disparate spatiotemporal scales, from fluid to kinetic. Amongst others, one of the most natural examples is plasma turbulence, which could be defined as a (fascinating) ``multi-scale disorder''~\citep{SchekochihinPPCF2008}.
These two processes, turbulence and MR, are indeed intimately related to each other, one process possibly feeding on the other and vice versa~\citep[e.g.,][and references therein]{ServidioNPG2011,ZhdankinAPJ2013,CerriCalifanoNJP2017,FranciAPJL2017,PucciAPJ2017,ComissoSironiPRL2018,MunozBuechnerPRE2018}.
In general, the intrinsically nonlinear, multi-scale nature of collisionless space plasma dynamics requires a numerical approach to the solution of these problems. This highlights the need for a compromise between the complexity of an appropriate theoretical framework and the available computational resources.

Numerical studies are based first on the choice of the plasma model used to describe, at best, the plasma dynamics of interest while keeping the computational cost of a simulation under control. 
The full-kinetic approach, based on the solution of the Vlasov equation for all plasma species (\textit{viz.}, ions, and electrons), represents the best mathematical model for the study of the dynamics of a collisionless, magnetized plasma at kinetic scale.
However, full-kinetic simulations usually come at a very large computational cost, too large to afford realistic plasma parameters and/or to include a wide enough range of spatiotemporal scales. 
On the opposite limit in terms of completeness of the plasma description and, as a consequence, of the computational cost of their numerical solution, one has ``fluid'' models. In this context, the ideal-MHD theory represents the simplest case of an infinitely conducting, single-fluid plasma where all kinetic effects at any characteristic scale are ignored. 
In between these two extremes, with respect to the full Vlasov approach, there is a long list of hybrid models \citep{TronciCamporealePOP2015}.
The first hybrid models were proposed a very long time ago to address quasi-neutral plasma dynamics, and they rely on a kinetic description of the ions and a fluid description of the electrons \citep[e.g.,][]{ByersJCP1978,HewettNielsonJCP1978,WinskeSSR1985}.
The hybrid approach has garnered a great deal of attention over the past two decades because it represents a very good compromise between the physics to be described and the computational cost of the numerical simulations.  The capability of hybrid models to catch the correct physics across the ion kinetic scales and below, approaching the electron scales, depends on the model adopted for the electron response. The simplest case of mass-less isothermal electrons may be too simplified when applied to cases in which the electron response may have important feedback on the ion dynamics. In this respect, within the hybrid approach, it is possible to improve the electron response by first including finite electron inertia effects in Ohm's  law~\citep[e.g.,][]{ValentiniJCP2007,MunozCPC2018}.
A step in this direction was taken in \citet{Lee2016PoP}, where a hybrid model with a static and gyrotropic equation of state for a mass-less electron fluid was considered. However, that equation of state was derived considering only the adiabatic trapping of electrons as the primary source of anisotropy.
A further step can be taken by introducing a more general fluid model that includes dynamic electron-pressure anisotropy~\citep[][]{ChewRSPSA1956,HunanaJPlPh2019a}, electron finite-Larmor-radius (FLR) effects, and/or electron Landau damping~\citep[e.g.,][]{HammettPerkinsPRL1990,SulemPassotJPP2015,HunanaJPlPh2019b}.

In general, hybrid-kinetic models have proved to be capable of satisfactorily catching the main kinetic physics at play for a large number of problems, ranging from fluid and kinetic instabilities~\citep{Hellinger2000,Matteini2006,calimirror2008,HenriPOP2013,KunzPRL2014} to collisionless shocks~\citep[][]{Lembege2009,CaprioliSpitkovskyAPJL2013,WeidlPOP2016}, dynamo effects~\citep{RinconPNAS2016,StOngeApJ2018}, MR~\citep[][]{Lee2016PoP,PalmrothAnGeo2017,CerriCalifanoNJP2017,FranciAPJL2017,WangPOP2019,cal2020}, and kinetic-scale turbulence~\citep[][]{ServidioJPP2015,GroseljAPJ2017,CerriAPJL2018,CerriFrASS2019,HellingerAPJ2019,WangPOP2019}.
The main goal of our project is to investigate the possibility of improving the electron description in hybrid-kinetic models.
The starting point is the ``hybrid Vlasov--Maxwell'' (HVM) code \citep{MangeneyJCP2002,ValentiniJCP2007}, which is already equipped with a fluid model in which electrons are described as an isotropic, isothermal fluid with finite inertia. The HVM code has recently been upgraded in order to implement a more sophisticated model for the electron fluid. This includes evolution equations for the anisotropic (gyrotropic) electron pressures, $p_{\|,{\rm e}}$ and $p_{\perp,{\rm e}}$ (where $\|$ and $\perp$  refer to the local magnetic-field direction, $\bb{b}=\bb{B}/|\bb{B}|$), and a Landau-fluid (LF) closure for the parallel transport of the gyrotropic electron thermal energy along field lines (i.e., parallel heat fluxes, $q_{\|,{\rm e}}$ and $q_{\perp,{\rm e}}$). Hereafter, we refer to this new model as ``hybrid Vlasov--Landau-fluid'' (HVLF).
The idea is to include within a hybrid description the relevant electron pressure-anisotropy effects and a fluid model for the electron-kinetic response that still holds in a nonlinear regime.
Therefore, the LF model implemented in the HVLF code goes beyond the early attempts to include these effects in simplified settings~\citep[e.g.,][]{HammettPerkinsPRL1990,SnyderPhPl1997,PassotSulemPhPl2007} and is based on the approach presented by \citet{SulemPassotJPP2015}.

In this work we consider the problem of MR, focusing on its nonlinear evolution. 
Since MR is a typical process in which both ion- and electron-kinetic effects at sub-ion scales drive an energy conversion that can feedback into the ``large'' (fluid) scale of the system, it represents a suitable problem in which to employ the HVLF model.
In fact, as highlighted by both simulations and observations, the heating associated with this process is typically anisotropic, exhibiting an enhancement of the parallel heating near the region where electrons are demagnetized, the so-called electron diffusion region \citep[EDR; see, e.g.,][]{ChenJGRA2008,DaughtonPhPl2007}. Magnetic reconnection naturally drives pressure anisotropies, which in turn constitute a free-energy source for secondary instabilities, most notably fire-hose instabilities (FHIs) and mirror instabilities (MIs). Their development bounds the plasma distribution in parameter space during the nonlinear stage of MR, as also seems to be the case in the turbulent solar wind \citep[SW; e.g.,][]{HellingerGeoRL2006,MatteiniGRL2007,BalePRL2009}. 
Here an MR simulation performed with the HVLF model is compared with equivalent simulations  (i.e., identical initial setups) performed with (i) the standard HVM code with isothermal electrons and finite electron-inertia effects and (ii) a fully kinetic model employing the semi-implicit particle-in-cell ``iPic3D'' code \citep{markidis2010multi,lapenta2012particle}. 

In Sect. \ref{sec:models} we introduce the three different models that have been used in this work and the initial simulation setup.
Numerical results are then detailed in Sect. \ref{sec:results}, starting with an analysis of the reconnection dynamics in Sect. \ref{subsec:dynamics} (e.g., reconnected flux and reconnection rate, as well as the properties of the relevant fields near the EDR). In Sect. \ref{subsec:anisotropy} we focus on the properties (and possible role) of the pressure anisotropy that is developed by protons and electrons during the nonlinear stage of MR. 
In Sect. \ref{sec:discussion} we summarize the results and outline our conclusions.

\section{Kinetic plasma models and MR setup}\label{sec:models}

In this section, we present the three different kinetic models that have been used in this work. 
These are (i) a hybrid-kinetic model, where the electrons are approximated either (a) as isotropic and isothermal fluid or (b) as anisotropic fluid with an LF closure, and (ii) a full-kinetic model.
It should be noted that the hybrid-kinetic approximation also keeps finite-electron-mass effects in both cases.

It is important to point out that, while classic hybrid-kinetic models exclude electron kinetic effects, such as electron Landau damping~\citep[see, e.g.,][]{ToldNJP2016b,CamporealeBurgessJPP2017}, this downside is partially cured when an LF type of closure is adopted for the electron fluid~\citep[e.g.,][]{HammettPerkinsPRL1990,SulemPassotJPP2015}.

\subsection{Hybrid-kinetic models}\label{subsec:HKmodels}

The hybrid-kinetic approximations consists of a quasi-neutral plasma, $n_\mathrm{p}=n_\mathrm{e}\equiv n$, where fully kinetic ions (protons in this case) are coupled to a fluid-electron model~\citep[][]{WinskeSSR1985}.
The evolution of the proton distribution function, $f_\mathrm{p}(\bb{x},\bb{v},t)$, follows the Vlasov equation,
\begin{equation}\label{eq:Vlasov}
\pD{t}{f_\mathrm{p}}\, +\, \bb{v}\bcdot\grad f_\mathrm{p}\, +\, \frac{e}{m_\mathrm{p}}\left(\bb{E} + \frac{\bb{v}}{c}\btimes\bb{B}\right)\bcdot \pD{\bb{v}}{f_\mathrm{p}}\, =\, 0\,,
\end{equation}
while the magnetic field, $\bb{B}$, evolves following Faraday's law of induction,
\begin{equation}\label{eq:Faraday}
\pD{t}{\bb{B}}\, =\, -c\,\grad\btimes\bb{E}\,,
\end{equation}
where $e$ and $m_{\rm p}$ are the proton electric charge and mass, respectively, while $c$ is the speed of light.
The proton dynamics is coupled to the electron fluid through a generalized Ohm's law that provides the electric field, $\bb{E}$,
\begin{equation}\label{eq:Ohm}
\bigl(1-d_\mathrm{e}^2\nabla^2\bigr)\bb{E}\, =\, 
-\frac{\bb{u}_\mathrm{e}\btimes\bb{B}}{c}\, 
+\frac{\grad\bcdot\bs{\Pi}_\mathrm{e}}{en}\, 
+ d_\mathrm{e}^2\grad\bcdot\bigl[n(\bb{u}_\mathrm{p}\bb{u}_\mathrm{p}-\bb{u}_\mathrm{e}\bb{u}_\mathrm{e})\bigr]\,,
\end{equation}
where $d_\mathrm{e}$ is the electron inertial length, the corresponding terms in \eqref{eq:Ohm} taking into account finite-electron-mass effects, while $\bb{u}_{\rm p}$ and $\bb{u}_\mathrm{e}\equiv\bb{u}_\mathrm{p}-\bb{J}/en$ are the proton and the electron fluid velocities, respectively.
The current density $\bb{J}$ is computed from the magnetic field via the Amp\`ere's law, $4\pi\bb{J}/c=\grad\btimes\bb{B}$, where the displacement current has been neglected in the hybrid approximation, while $n$ and $\bb{u}_\mathrm{p}$ are obtained as velocity moments of the proton distribution function.

The set of equations \eqref{eq:Vlasov}--\eqref{eq:Ohm} needs to be closed by one (or more) equation(s) for the pressure tensor of the electrons, $\bs{\Pi}_\mathrm{e}$. Two different closures for the electron fluid are considered here, isothermal or LF electrons, and the corresponding equations are detailed in the following subsections.

For the numerical solution of \eqref{eq:Vlasov}--\eqref{eq:Ohm}, we adopt an upgraded version of the Eulerian HVM code~\citep{ValentiniJCP2007}, which now allows different electron closures. 
When a standard isothermal electron fluid is considered, we refer to it simply as HVM (Sect.~\ref{subsubsec:HVM}), while when the new LF electron model is adopted, we refer to it as HVLF (Sect.~\ref{subsubsec:HVLF}). 
In the HVM code, an Eulerian approach is employed to numerically solve the Vlasov equation \eqref{eq:Vlasov} for the proton distribution function (i.e., $f_\mathrm{p}(\bb{x},\bb{v},t)$ is computed at each time step on a fixed, uniform grid sampling the simulated phase space $(\bb{x},\bb{v})$). This approach has the merit of avoiding any statistical noise that is typical of particle-in-cell (PIC) methods, but at the price of a large computational cost and memory requirements (when compared, for instance, to typical state-of-the-art hybrid-PIC simulations with similar parameters~\citep[e.g.,][]{CerriJPP2017}). 
To advance in time the proton distribution function, the HVM code implements an extended version of Knorr's splitting scheme that accounts for the electromagnetic case~\citep{MangeneyJCP2002}.
Furthermore, to efficiently couple the Vlasov equation for the proton distribution function with the Maxwell equations, the HVM code also employs the so-called current advancement method (CAM) \citep{CAM}. 
In the CAM method, in order to advance the distribution function from $t$ to $t+\Delta t$, electromagnetic fields at the half-step $t+\Delta t/2$ are needed. These fields depend upon the moments of the distribution function via the generalized Ohm's law \eqref{eq:Ohm}, so that in principle $f_{\rm p}(t+\Delta t/2)$ would also be needed. However, at this step of the algorithm, only $f_{\rm p}(t)$ and  its ``advection-only advancement''\footnote{Meaning that, due to the splitting scheme, $f_{\rm p}(t)$ has been advanced to $t+\Delta t/2$ using only the spatial-advection part of the Vlasov equation, $\partial f_{\rm p}/\partial_t+\bb{v}\bcdot\bb{\nabla}f_{\rm p}=0$, and not yet its electromagnetic part to also advance $f_{\rm p}$  in velocity space.} are known from the splitting scheme. 
The CAM approach can use these known states to provide an approximation of the electromagnetic fields, which is sufficiently good in the sense that it does not lower the whole algorithm accuracy (which is second-order in time). Further details about the HVM numerical scheme can be found in \citet{ValentiniJCP2007} and references therein.

\subsubsection{HVM: Isothermal electron fluid}\label{subsubsec:HVM}

The simplest class of closures is represented by a isotropic, barotropic fluid, that is, where the pressure tensor is isotropic, $\bs{\Pi}_\mathrm{e}=p_\mathrm{e}{\bf I}$ (${\bf I}$ being the identity tensor), and the scalar pressure is a function of the density only, $p_{\mathrm{e}}=p_{\mathrm{e}}(n_{\mathrm{e}})$. 
The first closure considered in our hybrid model is represented by isothermal electrons:
\begin{equation}
    \bs{\Pi}_\mathrm{e} = p_{\mathrm{e}}\bs{\mathrm{I}}\,,\quad p_{\mathrm{e}}=T_{0,\mathrm{e}}n_{\mathrm{e}}\,,
\end{equation}
where $T_{0,\mathrm{e}}$ is the constant, homogeneous electron temperature.

\subsubsection{HVLF: Landau-fluid electrons}\label{subsubsec:HVLF}

In magnetized, collisionless plasmas, the isotropy condition is rarely met. 
Charged particles indeed freely stream along $\bb{B}$, while they gyrate around the field lines. 
It is therefore natural to distinguish between the parallel and perpendicular response of the plasma, especially for what concerns the distribution of energy between the different degrees of freedom. 
To the lowest (zeroth) order in an FLR expansion of the pressure tensor, this can be taken into account by a ``gyrotropic'' pressure \citep{ChewRSPSA1956},
\begin{equation}\label{eq:Pi_el_gyrotropic}
    \bs{\Pi}_\mathrm{e} = p_{\parallel,\mathrm{e}}\bb{bb} + p_{\perp,\mathrm{e}}(\bs{\mathrm{I}}-\bb{bb})\,,
\end{equation}
where $\bb{bb}$ is the dyadic tensor of the magnetic-field unit vector, $\bb{b}=\bb{B}/B$.
Then, we explicitly solve the dynamic equations for the evolution of $p_{\mathrm{e,\parallel}}$ and $p_{\mathrm{e,\perp}}$. These are obtained via integration over $\bb{v}$ of Eq.~\eqref{eq:Vlasov} (but for electrons) multiplied by the dyadic $(\bb{v}-\bb{u}_{\rm{e}})(\bb{v}-\bb{u}_{\rm{e}})$, under the assumption of gyrotropic pressure \citep[see][and references therein]{HunanaJPlPh2019a}, and they read
\begin{subequations}
    \begin{equation}\label{eq:dppara_dt_el}
        \frac{\mathrm{d}p_{\parallel,\mathrm{e}}}{\mathrm{d}t} = - p_{\parallel,\mathrm{e}}\grad\bcdot\bb{u}_{\mathrm{e}} - 2p_{\parallel,\mathrm{e}}\bb{b}\bcdot\grad\bb{u}_{\mathrm{e}}\bcdot\bb{b} - \grad\bcdot(q_{\parallel,\mathrm{e}}\bb{b}) + 2q_{\perp,\mathrm{e}}\grad\bcdot\bb{b}\,,
    \end{equation}
    \begin{align}\label{eq:dpperp_dt_el}
        \frac{\mathrm{d}p_{\perp,\mathrm{e}}}{\mathrm{d}t} = &- 2p_{\perp,\mathrm{e}}\grad\bcdot\bb{u}_{\mathrm{e}} +\\&+ p_{\perp,\mathrm{e}}\bb{b}\bcdot\grad\bb{u}_{\mathrm{e}}\bcdot\bb{b} - \grad\bcdot(q_{\perp,\mathrm{e}}\bb{b}) - q_{\perp,\mathrm{e}}\grad\bcdot\bb{b}\,,\notag
    \end{align}
\end{subequations}
where $q_{\parallel,\mathrm{e}}$ and $q_{\perp,\mathrm{e}}$ are the two scalars determining the parallel component of a gyrotropic heat-flux tensor (i.e., the flux of parallel and perpendicular thermal energy along the magnetic lines, respectively). In the hybrid-Vlasov code, these equations are used to advance in time $p_{\|,{\rm e}}$ and $p_{\perp,{\rm e}}$ employing a third-order Runge-Kutta method. Also, a sub-stepping approach is implemented for these equations (meaning that, while the proton distribution function is advanced from $t$ to $t+\Delta t$ by means of a single time step of ``length'' $\Delta t$, the electrons pressure is simultaneously advanced with $n_{\text{sub}}$ sub-steps, each of length $\Delta t/n_{\text{sub}}$). We remark here that this work is focused on the physics that the HVLF model can reproduce when compared to a full-kinetic or an HVM approach (in the context of standard 2D MR), whereas a detailed technical paper focusing on the numerical method is currently in preparation.

At this point, we still need a closure for $q_{\parallel,\mathrm{e}}$ and $q_{\perp,\mathrm{e}}$. 
For instance, in the Chew–Goldberger–Low \citep[CGL,][]{ChewRSPSA1956} double-adiabatic model they are set to zero (hence the name).
Here, we adopt an LF type of closure for the electron heat-flux coefficients, that is, a closure that provides an approximated linear response based on kinetic theory. This closure was presented in \citet{SulemPassotJPP2015}, and its derivation consists in substituting the plasma response function by three-pole and one-pole Padé approximants in the formulas for the heat fluxes and the temperatures, respectively, which in turn are given by the linear kinetic theory. The closure is then expressed as closed formulas for the heat-flux density coefficients, which in the case of the electrons are
\begin{subequations}\label{subeqs:qpara_qperp_el}
    \begin{equation}\label{eq:qpara_el}
        q_{\parallel,\mathrm{e}} = - \overline{p}_{\parallel,\mathrm{e}}v_{th,\parallel,\mathrm{e}}{\sqrt{\frac{8}{\pi}}}\mathcal{H}\left[\frac{T_{\parallel,\mathrm{e}}}{\overline{T}_{\parallel,\mathrm{e}}}\right]\,,
    \end{equation}
    \begin{equation}\label{eq:qperp_el}
        \begin{split}
            q_{\perp,\mathrm{e}} =& - \frac{\overline{p}_{\perp,\mathrm{e}}(\overline{T}_{\perp,\mathrm{e}}-\overline{T}_{\parallel,\mathrm{e}})}{\Omega_{\rm p} m_{\rm p}}\bb{b}\bcdot\left(\grad\btimes\frac{\bb{B}}{\overline{B}}\right) -\\&- \overline{p}_{\perp,\mathrm{e}}v_{th,\parallel,\mathrm{e}}\sqrt{\frac{2}{\pi}}\mathcal{H}\left[\frac{T_{\perp,\mathrm{e}}}{\overline{T}_{\perp,\mathrm{e}}} + \left(\frac{\overline{T}_{\perp,\mathrm{e}}}{\overline{T}_{\parallel,\mathrm{e}}} - 1\right)\frac{B}{\overline{B}}\right]\,.
        \end{split}
    \end{equation}
\end{subequations}
Here $T_{\| \mathrm{e}}$ and $T_{\perp \mathrm{e}}$ denote the parallel and perpendicular electron temperatures, $v_{th,\parallel,\mathrm{e}}=\sqrt{T_{\| 0, \mathrm{e}}/m_{\rm e}}$ is the electron parallel thermal velocity of the unperturbed plasma (with $m_{\rm e}$ the electron mass), and $\Omega_{\rm p}=eB_0/m_{\rm p}c$ is the proton cyclotron frequency (with $B_0$ the intensity of the magnetic field in the unperturbed state).
In Equations \eqref{subeqs:qpara_qperp_el} the over-bar, $\overline{(\dots)}$, denotes the space average and $\mathcal{H}$ is the operator represented, in Fourier space, by
\begin{equation}\label{eq:H}
    \widehat{\mathcal{H}[g]}(\bb{k}) = \frac{\widehat{(\bb{b}\bcdot\grad g)}(\bb{k})}{\sqrt{\bb{k}\bcdot\overline{\bb{bb}}\bcdot\bb{k}}}\,,
\end{equation}
where the hat, $\widehat{(\dots)}$, denotes Fourier-transformed quantities. This operator effectively takes into account the distortion of magnetic-field lines in the nonlinear regime.
This operator should have been the (negative) Hilbert operator along perturbed field lines going through each point,  as in Eq.~(50) of \citet{SnyderPhPl1997}. This however involves non-affordable computations in large simulations.
Therefore, we adopted the semi-phenomenological modeling of $\mathcal{H}$, still local in Fourier space, presented by \citet{PassotEPJD2014}.
In the linear regime, when $\bb{b}=\bb{e}_z$, the Fourier representation of $\mathcal{H}$ given by Eq.~\eqref{eq:H} reduces to $i\,\mathrm{sgn}(k_z)=i\,k_z/|k_z|$, as originally introduced by \citet{HammettPerkinsPRL1990}.
This semi-phenomenological operator is formally a good approximation only where the magnetic field is not strongly curved, a condition that could be challenged by a nonhomogeneous equilibrium (as in the case of this work).
We also stress that the LF closure defined through Eq. \eqref{eq:H} breaks down in the presence of magnetic nulls, since $\bb{b}$ cannot be defined where the magnetic field vanishes. In fact, the $\bb{b}\bcdot\bb{\nabla}$ operator becomes ill-behaved whenever $|\bb{B}|$ approaches zero. For this reason, in this work, we adopt a finite value for the initial guide field (low enough not to slow down the reconnection dynamics, but sufficiently large to avoid the emergence of magnetic nulls throughout the simulation).

Additionally, we mention that a possible approach could be to add also some artificial constraints to prevent runaway anomalies. This would be the case when hard-coded ``anisotropy limiters'' are employed to bound the values of the pressure anisotropy within a physically ``sensible'' range, something usually adopted when modeling (nearly) collisionless plasmas as fluids.
For example, to mimic microphysical instabilities that ``feed'' on the pressure anisotropy, limiters were added to collisionless MHD-like models to perform numerical studies of magneto-rotational instability \citep{SharmaApJ2006}, to perform numerical studies of the turbulent dynamo in the intracluster medium \citep{Santos-LimaApJ2014,StOngeJPP2020}, and to derive a heating mechanism able to stably balance radiative cooling in the central regions of cold-core clusters of galaxies \citep{KunzMNRAS2011a}.
In all these examples the inclusion (via the limiters) of microphysical effects significantly changes the saturation level of macroscopic quantities.
However, the current work seems to suggest that, in the present application, the implemented LF closure alone performs quite adequately to that end (as long as the conditions for the LF model to be valid are satisfied). From the figures presented in Sect. 3.2, one can see that the electron anisotropy remains always bounded within a reasonable range. Nevertheless, we do not exclude that including such limiters for a certain class of problems could be an interesting option to be further investigated in the future.

Finally, we note that in this model we do not include electrons' FLR corrections to the gyrotropic pressure tensor in \eqref{eq:Pi_el_gyrotropic}. Therefore, although retaining a model for the electron Landau damping, we do not capture other kinetic effects at $k_\perp\rho_\mathrm{e}\sim1$, where $\rho_{\rm{e}}$ is the electron Larmor radius (see, e.g., \citet{SulemPassotJPP2015} for a detailed discussion). 
On the other hand, we stress that the electron model adopted here includes several additional effects with respect to the electron response adopted in a similar study by \citet{Lee2016PoP}. In fact, while their model only considers a static, gyrotropic equation of state ({\em viz.}, $p_{\parallel,\mathrm{e}}=p_{\parallel,\mathrm{e}}(n,B)$ and $p_{\perp,\mathrm{e}}=p_{\perp,\mathrm{e}}(n,B)$), our HVLF model includes dynamic equations \eqref{eq:dppara_dt_el}--\eqref{eq:dpperp_dt_el} for the gyrotropic electron-pressure components, LF closures \eqref{eq:qpara_el}--\eqref{eq:qperp_el} for the gyrotropic electron heat fluxes, and finite electron-inertia effect in the generalized Ohm's law \eqref{eq:Ohm}.

\subsection{Full-kinetic model: iPIC}\label{subsec:FKmodel}

Compared to the LF closure implemented in the HVLF run, a full-kinetic model will provide also the non-gyrotropic effects of the dynamics of the electrons (i.e., those linked with FLR physics). Also, the quasi-neutrality ansatz will be removed, meaning that charge separation is not forced to be zero.

The most widely used method to numerically describe the kinetic dynamics of a plasma system is the PIC method. The majority of PIC codes use an explicit method to advance the electric and magnetic fields in time \citep{bowers2008ultrahigh}. The explicitness of the method introduces strong restrictions regarding the length and timescales to resolve in the numerical integration: (i) the plasma Debye length, the shortest length scale in the system that represents the characteristic screening scale length of a single charge by collective plasma motions, must be resolved by the spatial grid; (ii) the electron plasma frequency has to be resolved temporally; (iii) a Courant condition on the speed of light limits even further the grid size. Implicit and semi-implicit PIC methods have been developed to reduce such constraints. They are not subject to the aforementioned restrictions and allows for the simulation of fully kinetic plasma systems in a range that goes from fluid scale to a fraction of the electron skin depth.

Considering the size and the resolution of the system presented here, we opted for the use of the semi-implicit method implemented on the code ``iPic3D'' \citep{markidis2010multi,lapenta2012particle}. The semi-implicitness of the method arises from an approximated calculation of the source terms (currents and density) at the iterative step ``n+1,'' which are computed by a Taylor expansion in time of the same quantities at the step ``n.'' As explained in \citet{markidis2010multi}, the semi-implicit method has a reduced computational cost as compared with the implicit method and, clearly, with every explicit method. As a drawback, the semi-implicit method requires the following stability condition to be satisfied: $v_{th,e}\Delta t/\Delta x < 1$, where $v_{th,e}$ is the electron thermal speed, $\Delta t$ is the time step, and $\Delta x$ is the grid step. The latter is always less restrictive than the Courant condition on the speed of light $c$ for the explicit methods, which reads $c \Delta t/\Delta x < 1$.
The code iPic3D solves the full Vlasov-Maxwell set of equations for multiple species, in this case, protons and electrons. The physics resolved in a particular simulation depends on the system resolution and on the time step. Contrary to explicit methods where all relevant spatial and temporal scales are resolved, the semi-implicit method can resolve phenomena that have a spatial scale larger than the grid size and a frequency higher than the inverse of the time step. 
Thus, the presence or absence of phenomena such as quasi-neutrality violation depends on the numerical resolution.
In this study, the implicit nature of the code is effective in removing any wave generated by charge separation, so that the quasi-neutrality is mostly preserved.

\subsection{Simulation setup}\label{subsec:setup}

For each of the plasma models mentioned above, we performed a numerical integration using the same initial condition on a 2D spatial domain of size $L_x = 24\pi d_{\rm p}$ and $L_y = 12\pi d_{\rm p}$, discretized with $N_x\times N_y = 1024\times 512$ grid points.
The velocity space domain in the HVM and HVLF hybrid simulations is bounded by $|v|\leq 6.4v_{\rm th,p}$ in each velocity direction, where $v_{\rm th,p}\equiv\sqrt{T_{\rm 0,p}/m_{\rm p}}$ is the proton thermal velocity associated with the initial proton temperature $T_{\rm 0,p}$. 
The velocity grid has been discretized by $51^3$ grid points.
The iPIC (fully kinetic) run employs $14000$ particle per cell (PPC), $7000$ for each species. The hybrid simulations also include finite electron-inertia effects, and in all three models we adopted a reduced mass ratio of $m_{\rm p}/m_{\rm e}=100$.

The initial setup consists of a ``double-Harris (DH) sheet''  configuration~\citep{HarrisNCim1962} to accommodate for periodic boundary conditions:
\begin{align}\label{eq:B_equilinrium}
    B_x & = B_0\left[\tanh\left(\frac{y-y_1}{L_1}\right)-\tanh\left(\frac{y-y_2}{L_2}\right)-1\right]\;,\notag\\
    B_y & = 0\;,\\
    B_z & = B_0/4\;,\notag
\end{align}
where $L_1=0.85d_{\rm p}$ and $L_2=1.7d_{\rm p}$ are the two shear widths, $y_1=L_1 L_y/[2(L_1+L_2)]$ and $y_2=L_y-(L_2 L_y)/[2(L_1+L_2)]$ their positions, and $B_0=1$ is a constant value used for normalization.
Unlike the original Harris' equilibrium, we also included a ``weak'' guide field in order to have a non-vanishing magnetic field to allow for the LF closure. In other words, if the magnetic field vanishes, equations for parallel and perpendicular quantities become unjustifiable.
We set the guide field value to a quarter of the (asymptotic) reconnecting field $B_x$, so as not to slow down reconnection significantly~\citep[see, e.g.,][]{ricci2004collisionless,DaughtonJGRA2005,ShiarXiv2020}.

\begin{figure*}[ht!]
    \centering%
    \includegraphics[width=1.0\textwidth]{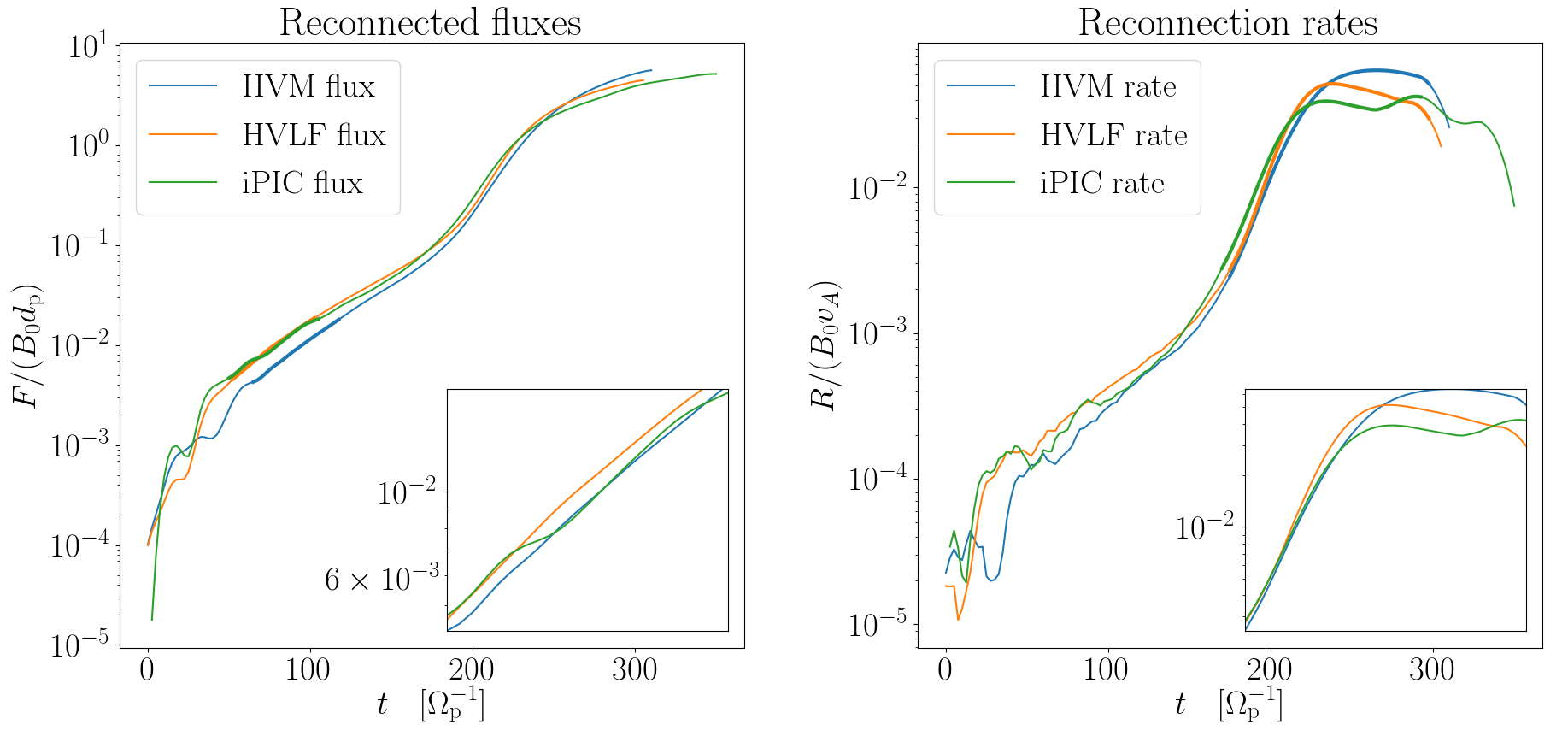}%
    \caption{Reconnected flux, $F$ (left), and reconnection rate, $R$ (right), for the main X point in the simulations. Within each panel, an inset shows a zoom-in on a specific stage of these curves (thickened on the main plot), shifted in time, for the sake of comparison.}
    \label{fig:rec_flux_rate}
\end{figure*}

The particle distribution function of the species ${\rm a}={\rm p,e}$ is initialized consistently with the magnetic-field configuration in Eq. \eqref{eq:B_equilinrium}:
\begin{align}\label{eq:f_eqilibrium}
    f^{DH}_{\mathrm{a}}(y,v_x,v_y,v_z)&=\frac{1}{(2\pi v_{\mathrm{th,\mathrm{a}}})^3}
    \left[n_1(y)\exp{\left(-\frac{v_x^2+v_y^2+(v_z-V_{1,\mathrm{a}})^2}{2v_{\mathrm{th,a}}^2}\right)}+\right.\notag\\
    &+n_2(y)\exp\left(-\frac{v_x^2+v_y^2+(v_z-V_{2,\mathrm{a}})^2}{2v_{\mathrm{th,a}}^2}\right)+\\
    &+\left.n_{\mathrm{b}}\exp\left(-\frac{v_x^2+v_y^2+v_z^2}{2v_{\mathrm{th,a}}^2}\right)\right]\;,\notag
\end{align}
where the density profiles of the ``background'' and sheets populations, namely $n_{\rm b}$, $n_1$, and $n_2$, respectively, are given by
\begin{align}\label{eq:n_equilibrium}
    n_1(y) &= \frac{B_0^2}{2(T_{0,\mathrm{p}}+T_{0,\mathrm{e}})\cosh^2\left(\frac{y-y_1}{L_1}\right)}\;,\notag\\
    n_2(y) &= \frac{B_0^2}{2(T_{0,\mathrm{p}}+T_{0,\mathrm{e}})\cosh^2\left(\frac{y-y_2}{L_2}\right)}\;,\\
    n_{\mathrm{b}} &= n_0\;,\notag
\end{align}
and $n_0=1$ is a constant value used for normalization (also, in our units the Boltzmann constant is $\kappa_{_B}=1$). 
Here $m_\mathrm{a}$ and $T_{0,\mathrm{a}}$ are the mass and the initial (uniform) temperature of the species, $v_{\mathrm{th,a}}=\sqrt{T_{0,\mathrm{a}}/m_{\mathrm{a}}}$ its initial thermal velocity and $V_{1,\mathrm{a}}=-2T_{0,\mathrm{a}}/(B_0L_1)$ and $V_{2,\mathrm{a}}=2T_{0,\mathrm{a}}/(B_0L_2)$ the velocities related to the current sheets.
If one considers a single current sheet, $L_2\to+\infty$, then Eqs. \eqref{eq:B_equilinrium}, \eqref{eq:f_eqilibrium}, and \eqref{eq:n_equilibrium} form a consistent kinetic equilibrium \citep[see][]{AllansonGeoRL2017}. Due to the nonlinear nature of the Vlasov-Maxwell system of equations, the superposition of two Harris' sheets is no longer a Vlasov equilibrium. Nevertheless, with our choice of parameters the two sheets are well separated in space: at the peak of the first current sheet ($n_1\simeq n_0$), the second one is negligible ($n_2\sim10^{-10}n_0$), and vice versa.
Moreover, the choice of $L_2 = 2L_1$ makes the second current sheet quiescent over the entire simulation time (i.e., no reconnection nor other instabilities develop).

For our simulations, we considered\footnote{To be precise, this $\beta_{\mathrm{p}}$ is the in-plane beta given by $8\pi P_{\mathrm{p}}/B_x^2=1$.} $\beta_{\mathrm{p}}=1$, 
$T_{0,\mathrm{e}}/T_{0,\mathrm{p}}=1/4$, and $T_{0,\parallel,\mathrm{a}}=T_{0,\perp,\mathrm{a}}$. The choice of adopting colder electrons follows from the absence of electron-FLR effects in the HVLF model. At the same time, the hybrid models employed in this work do include electron-inertia effects associated with $d_e$. Therefore, in order to resolve the electron inertial length $d_{\rm e}$ while keeping the electron Larmor radius below the grid resolution, we adopted $\beta_{\rm e}<1$. The initial equilibrium is perturbed by long wavelength, random phase magnetic-field fluctuations with $1\leq kd_{\rm p}\leq 9$, where $k\equiv(k_x^2+k_y^2)^{1/2}$ (all simulations employ the same phases for the initial perturbations). 
By initializing a superposition of different modes, we do not impose a single X point to emerge at a given position. 
Furthermore, in order to mitigate the effect of PIC noise, the maximum amplitude $\epsilon\equiv\max(|\delta B|)$ of the initial perturbations used in hybrid simulations, namely $\epsilon_{\rm hyb}=0.01$, is increased by a factor of three in the iPIC case, $\epsilon_{\rm pic}=0.03$. Other characterizing parameters for the simulations are $\Delta x=\Delta y=0.074d_{\rm{p}}=0.74d_{\rm{e}}$, $\Delta t^{\text{(HVM)}}=0.005\Omega_{\rm p}^{-1}$ (with $3$ electrons sub-steps),
$\Delta t^{\text{(iPIC)}}=0.2\omega_{P,\rm{p}}^{-1}$, and for the iPIC simulation only, also $v_A/c=0.01$, $\omega_{P,\rm{e}}/\Omega_{\rm e}=10$, $\lambda_D/d_{\rm{e}}=0.035$.
Within our choice of parameters, in the asymptotic region of the Harris field the electron gyro-motion is characterized by $\Omega_{\rm e}=100\Omega_{\rm p}$ and $\rho_{\rm e} = 0.035d_{\rm p}$ (we recall that $m_{\rm p}/m_{\rm e}=100$).

Given $\Delta x$ and $\Delta t^{\text{(iPIC)}}$ defined above, the time and spatial (asymptotic) resolution of the iPIC simulation in terms of electrons' quantities is thus $\Delta t^{\text{(iPIC)}} = 0.2\Omega_{\rm e}^{-1}$ and $\Delta x = 2\rho_{\rm e}$ (corresponding to a frequency and wavenumber resolution of $\omega/\Omega_{\rm e}\lesssim 15.7$ and $k_\perp\rho_{\rm e}\lesssim1.6$, respectively). 
As a result, the electron gyration is not exactly resolved in space where the magnetic field is stronger (asymptotic region), but it is well-resolved in time everywhere, with at least $\approx 30$ time steps in the worst-case scenario (if $\tau_{\rm e}=2\pi/\Omega_{\rm e}$ is the gyro-period, then $\Delta t^{\text{(iPIC)}}=\tau_{\rm e}/(10\pi)$). Therefore, phenomena at the electron cyclotron frequency can be described within the whole simulation domain. On the other hand, a limitation exists on the range of spatial scales that are resolved: although finite-$k_\perp\rho_{\mathrm{e}}$ effects can be captured by the iPIC simulation, to a certain extent, it is true that these effects do not extend far below the electron-Larmor-radius scale (at least in the asymptotic region, where the magnetic field is stronger; electrons' gyration is better resolved within the Harris sheet, where the magnetic field strength drops by a factor of up to four with respect to the asymptotic region).

\section{Numerical results}\label{sec:results}


In this section, we present an analysis of MR simulations within the different physical models introduced in Sect.~\ref{sec:models} (namely HVM, HVLF, and iPIC). Our analysis focuses on the following aspects: (i) the overall reconnection dynamics, (ii) the morphology of the EDR and outflows, and (iii) the pressure-anisotropy production and regulation.

In particular, we highlight the capability of the HVLF simulation of capturing part of the physics\footnote{Namely, the linear kinetic response (an approximation of it), including the linear Landau damping, and also the ability to include ``non-double-adiabatic'' anisotropy in the model.} missing in the HVM one and playing an important role in the dynamics, in agreement with the full PIC approach. Of course, the richer electron physics visible within the full PIC model is missing in the HVLF simulation. Nevertheless, this does not prevent the hybrid model adopted in the HVLF run from describing the reconnection dynamics quite satisfactorily.

\subsection{Reconnection dynamics}\label{subsec:dynamics}

\paragraph{\textbf{Reconnection rate.}} We characterize the global reconnection dynamics by means of (i) the reconnected flux and the associated reconnection rate (in other words, the amount of magnetic flux advected through the X point) and (ii) the physical structures formed during the reconnection and their nonlinear evolution.
In 2D, the reconnected flux $F$ is obtained by calculating the difference between the values of the magnetic flux function $\Psi$ evaluated at the X point and at the adjacent O-point as follows \citep[see][]{YeatesPhPl2011}:
\begin{equation}\label{eq:reconnected_flux}
    F(t)=|\Psi(t,\text{X point})-\Psi(t,\text{O-point})|\;.
\end{equation}
The associated reconnection rate, $R$, is obtained as the time derivative of $F$:
\begin{equation}\label{eq:reconnection_rate}
    R(t)=\frac{\mathrm{d}F}{\mathrm{d}t}\;.
\end{equation}

Although it can be shown that $R$ is equivalent to the more commonly used parallel electric field evaluated at the X point, $E_{\parallel}(\text{X Point})$, we point out that in PIC simulations the $\Psi$ function is less affected by numerical noise than the electric field.
Moreover, it does not require time-averaging to smooth out the oscillations typically observed as a consequence of a slight readjustment to the initial setup, rather than to the reconnection process itself.
In addition, the evolution of the reconnected flux allows the linear and nonlinear phases of the reconnection process among the different runs to be synchronized (and, thus, to be properly compared).

In Fig.~\ref{fig:rec_flux_rate} we show the reconnected flux $F$ (left panel) and the associated reconnection rate $R$ (right panel) obtained from the different simulations. 
From these plots we distinguish three consecutive phases: the linear stage of the instability (up to $t\sim150\Omega_{\rm p}^{-1}$), a subsequent nonlinear ``super-exponential'' phase (up to $t\sim230\Omega_{\rm p}^{-1}$), and a final quasi-stationary fully nonlinear regime.
The first phase presents the usual exponential growth of the modes, the second one is characterized by some modes growing faster than $\exp(\gamma_{\mathrm{lin,k}}t)$ \citep[hence the name super-exponential or quasi-explosive, as defined in][and \citeauthor{AydemirPhFlB1992} \citeyear{AydemirPhFlB1992}]{OttavianiPhRvL1993}, the third by a quasi-stationary reconnection rate.

\begin{figure*}[ht!]
\includegraphics[width=1.\textwidth]{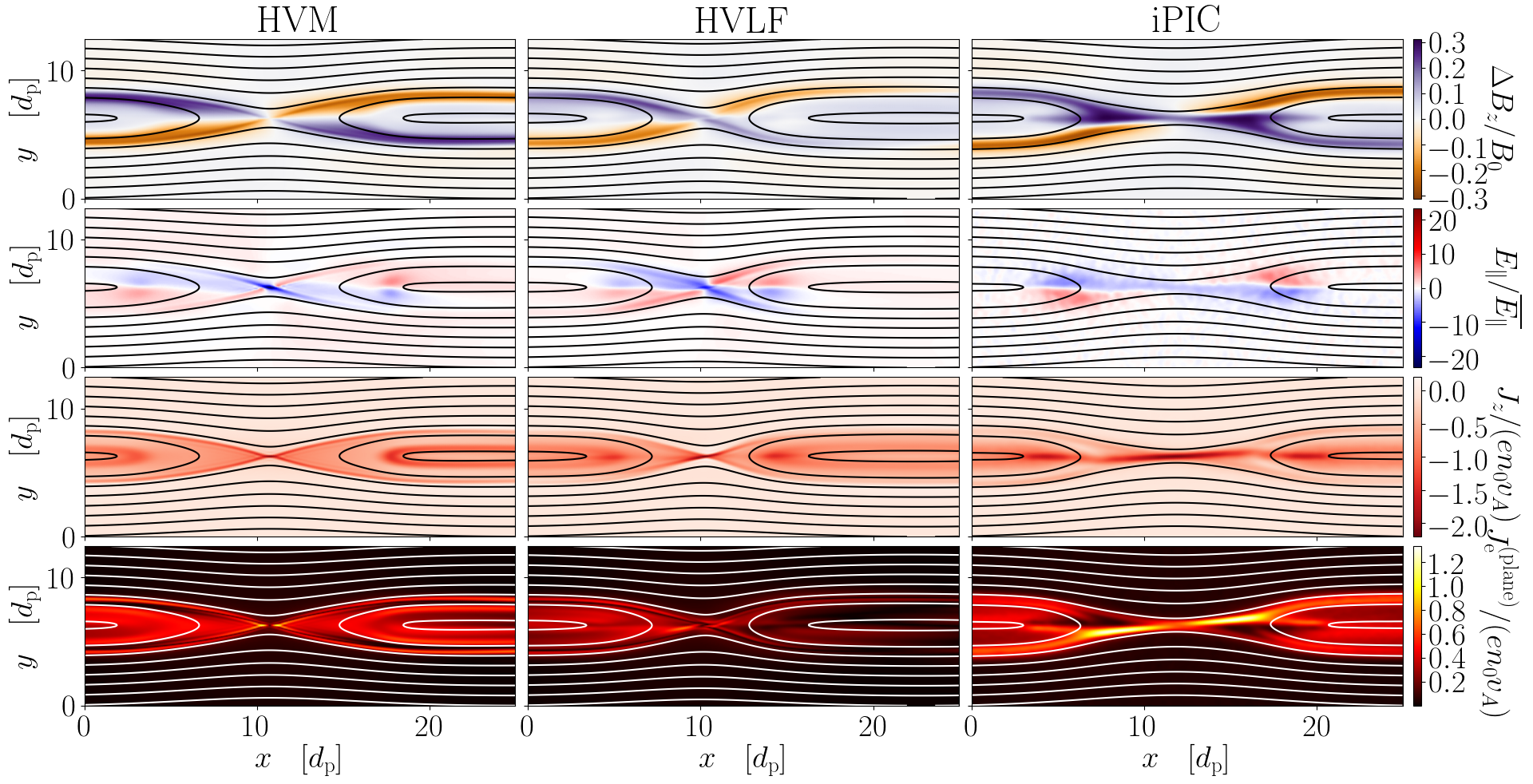}%
\caption{Heat maps of $\Delta B_z = B_z - B_z(t=0)$ (first row), $E_\|$ (second row), $J_z$ (third row), and $J_{\mathrm{e}}^{(\text{in-plane})}$ (fourth row) for the HVM (left column, $t=237.5\Omega_{\rm p}^{-1}$), HVLF (central column, $t=232.5\Omega_{\rm p}^{-1}$), and iPIC (right column, $t=235.0\Omega_{\rm p}^{-1}$) simulations. These times are chosen by fixing the value of the reconnected magnetic flux. This value is in the middle of the super-exponential phase. The plot is zoomed-in around the main X point. The $\overline{E_\|}$ is the r.m.s. value of $E_\|$ in the shown region.}%
\label{fig:2d_supexp_zoom}%
\end{figure*}

A qualitatively good agreement between the different simulations during the linear phase is observed, as shown in the left panel of Fig.~\ref{fig:rec_flux_rate} and its inset. However, a detailed, quantitative comparison of the modes' growth rates during the linear phase is limited by the uncertainties associated with finite-PPC noise.
Thus, we tested the convergence of the PIC simulation, in particular during the nonlinear stage, using different numbers of macro-particles, ranging from $800$ to $7000$ PPC per species (the higher number is the value used for the simulation presented here). This nonlinear phase of MR is the one of interest here. We focus on the late-time evolution when the process develops enough anisotropy to affect the dynamics, and thus the ``robustness'' of the electrons LF model can also be tested.

A qualitative agreement between the three simulations (see Fig.~\ref{fig:rec_flux_rate}, right panel and its inset) persists during the ``super-exponential'' phase, while they begin to differ during the quasi-stationary regime. This feature is the consequence of the different response times of the electron dynamics associated with the three different models.
During the quasi-steady stage, the (stationary) reconnection rate is higher for the HVM simulation (around a value $\sim 0.06$), while the HVLF one, which includes electron anisotropy, shows a better agreement with the fully kinetic case (both reaching a quasi-steady reconnection rate value around $\sim 0.04$).
The same trend for the quasi-stationary reconnection rates (i.e., higher with isotropic and isothermal electrons, lower in the full-kinetic case) was also reported in \cite{Lee2016PoP}.
Although the stationary reconnection rates are smaller than the ``usual'' value of about $0.1$~\citep[][]{CassakJPlPh2017}, it is worth noticing that this rate can indeed also depend on the background density~\citep[see][]{WuPhPl2011,DivinPhPl2019} or, more precisely, on the ratio between the background (i.e., asymptotic) and the peak (i.e., in the middle of the Harris sheet) values of the density.
In fact, most simulations adopt a background-to-peak density ratio of $n_{\rm{b}}/n_{\rm{H},0}=0.1-0.2$\footnote{Here, $n_{\rm{H},0}$ is the peak density value for the Harris's sheet. In our setup, we have $n_{\rm{H},0}=n_1(y_1)$.}, in order to better compare with early simulations setup~\citep[usually GEM-like configurations, see][]{BirnJGR2001} and with some magnetospheric data.
In \citet{CassakJPlPh2017} the so-called $0.1$ reconnection rate problem is addressed showing that indeed not all \emph{in situ} observations and/or simulations report exactly that value of the normalized reconnection rate. 
In our simulation, for instance, the background-to-peak density ratio is $n_{\rm{b}}/n_1(y_1)=1.25$ (cf. Eq. \eqref{eq:n_equilibrium}) and thus a lower stationary reconnection rate is consistent with previous works \citep[][]{DivinPhPl2019}.

\begin{figure*}
\includegraphics[width=1.\textwidth]{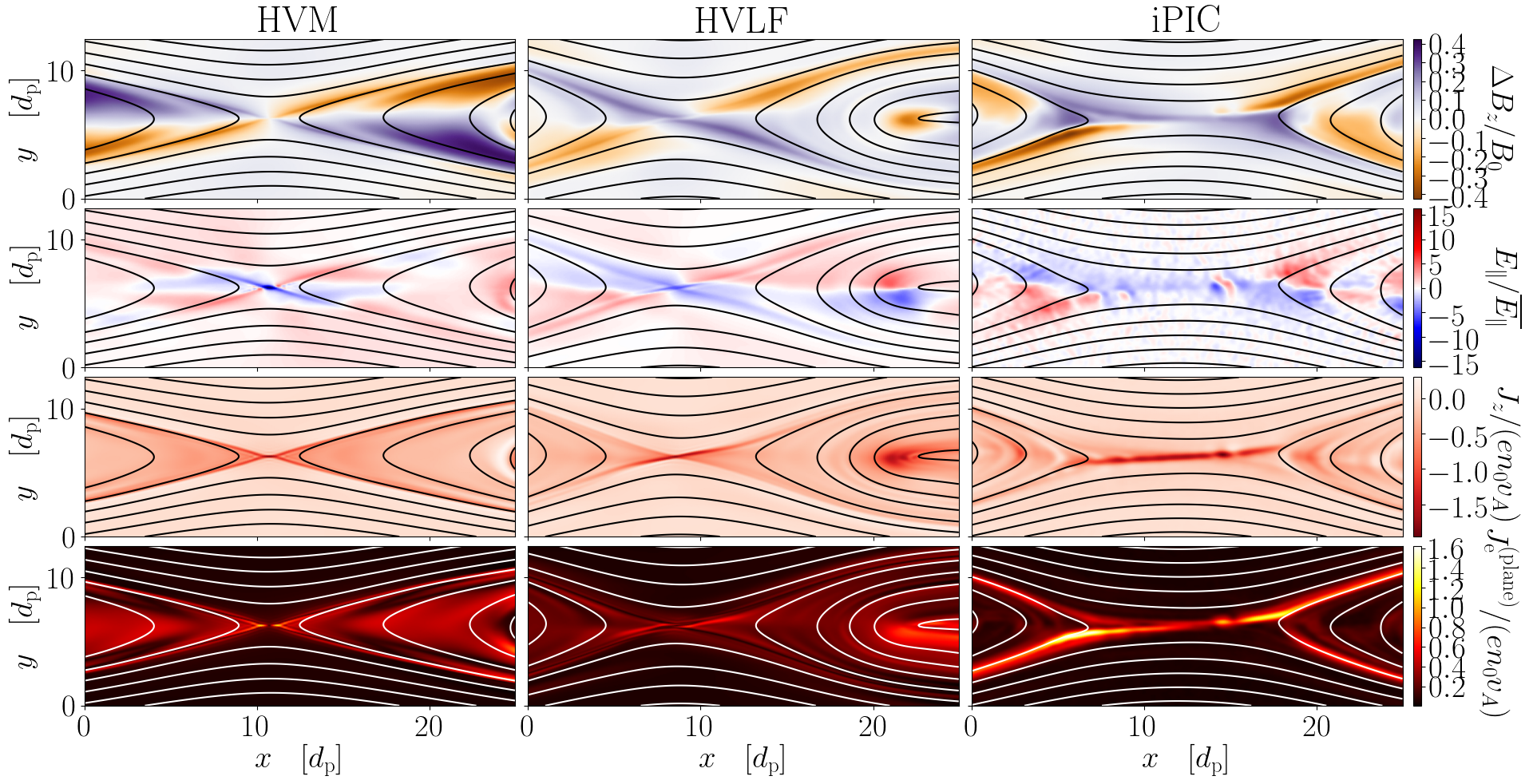}%
\caption{Same as Fig. \ref{fig:2d_supexp_zoom}, but the common value of reconnected flux is chosen in the nonlinear phase (for HVM at $t=287.5\Omega_{\rm p}^{-1}$, for HVLF at $t=305.0\Omega_{\rm p}^{-1}$, and for iPIC at $t=320.0\Omega_{\rm p}^{-1}$).}%
\label{fig:2d_nonlin_zoom}%
\end{figure*}

\paragraph{\textbf{Morphology.}} We now discuss the main features of the reconnection regions. In Fig.~\ref{fig:2d_supexp_zoom} and \ref{fig:2d_nonlin_zoom} we show for each simulation four fields of interest, namely $\Delta B_z=B_z-B_z(t=0)$, $E_\|$, $J_z$, and $J_{\mathrm{e}}^{(\text{in-plane})}$, in the region around the X point during the nonlinear phase. The two figures are taken before and during the stationary phase, respectively. 
If not otherwise stated when comparing results from different simulations, the time is chosen such that the corresponding value of the reconnected flux $F$ at that time is the same for all simulations (cf. Fig.~\ref{fig:rec_flux_rate}).
First of all, from Fig.~\ref{fig:2d_supexp_zoom} and \ref{fig:2d_nonlin_zoom} one observe that the morphology of the X point, as well as the symmetry of the various physical quantities around that region, highlights a significant difference between the three models, marking the distinction between the dynamics resulting from isothermal electrons (HVM) and from the electron's models adopted in the HVLF and iPIC simulations.
Moreover, as seen in the bottom row of these figures, in the HVLF and iPIC case one observes (more and more) elongated electron jets as compared to the HVM case. These jets, as well as the out-of-plane current, are tilted with respect to the direction of the neutral line. Another feature present in the HVLF and iPIC cases only is an asymmetric pattern in $B_z$ fluctuations within the EDR.
In the $E_\|$ field, more clearly seen in Fig. \ref{fig:2d_nonlin_zoom}, a similar structure is present in the three simulations. However, its typical extension is comparable in the HVLF and iPIC cases only, both quite larger than in the HVM one, during the quasi-stationary phase.
In the PIC simulation, as compared with the other fields, $E_\|$ is noisier (the reason why we prefer $\Psi$ to evaluate $F$) and ``fragmented.'' This fragmentation, also visible in other quantities, is due to the formation ad absorption of a plasmoid in the time range from $t\approx260.0\Omega_{\rm p}^{-1}$ to $t\approx290.0\Omega_{\rm p}^{-1}$ (cf. Fig.~\ref{fig:rec_flux_rate}). In this paper, we highlight other effects of this transient plasmoid; however, we can state that its formation did not pose a problem to the present study.

All the above-mentioned features have already been pointed out by \citet{EgedalPhPl2009} as typical signatures for fully kinetic models in the presence of a guide field. The same authors proposed a fluid closure for electrons \citep[see][]{LePhRvL2009}, which can be interpreted as an ``interpolation'' between an isothermal and a double-adiabatic closure, with the addition of kinetic features\footnote{Actually, it is derived from considerations about the distribution function, discriminating between trapped and passing electrons. Thus, it is more than a naive interpolation between two preexisting models. However, in the weak-trapping and strong-trapping limits it approaches the isothermal and double-adiabatic models, respectively.}. This closure has been validated in \citet{LePhRvL2009,LePhPl2010} and \citet{EgedalPhPl2013}, within a two-fluid framework, and an analogous configuration for the out-of-plane electrons current was obtained. However, the current sheet in the simulation with their model forms a smaller angle with the neutral line, if compared with an analogous full-kinetic run. In this sense, the HVLF results are closer to those observed in the full-kinetic iPIC case. 

\begin{figure*}[!ht]
\includegraphics[width=1.\textwidth]{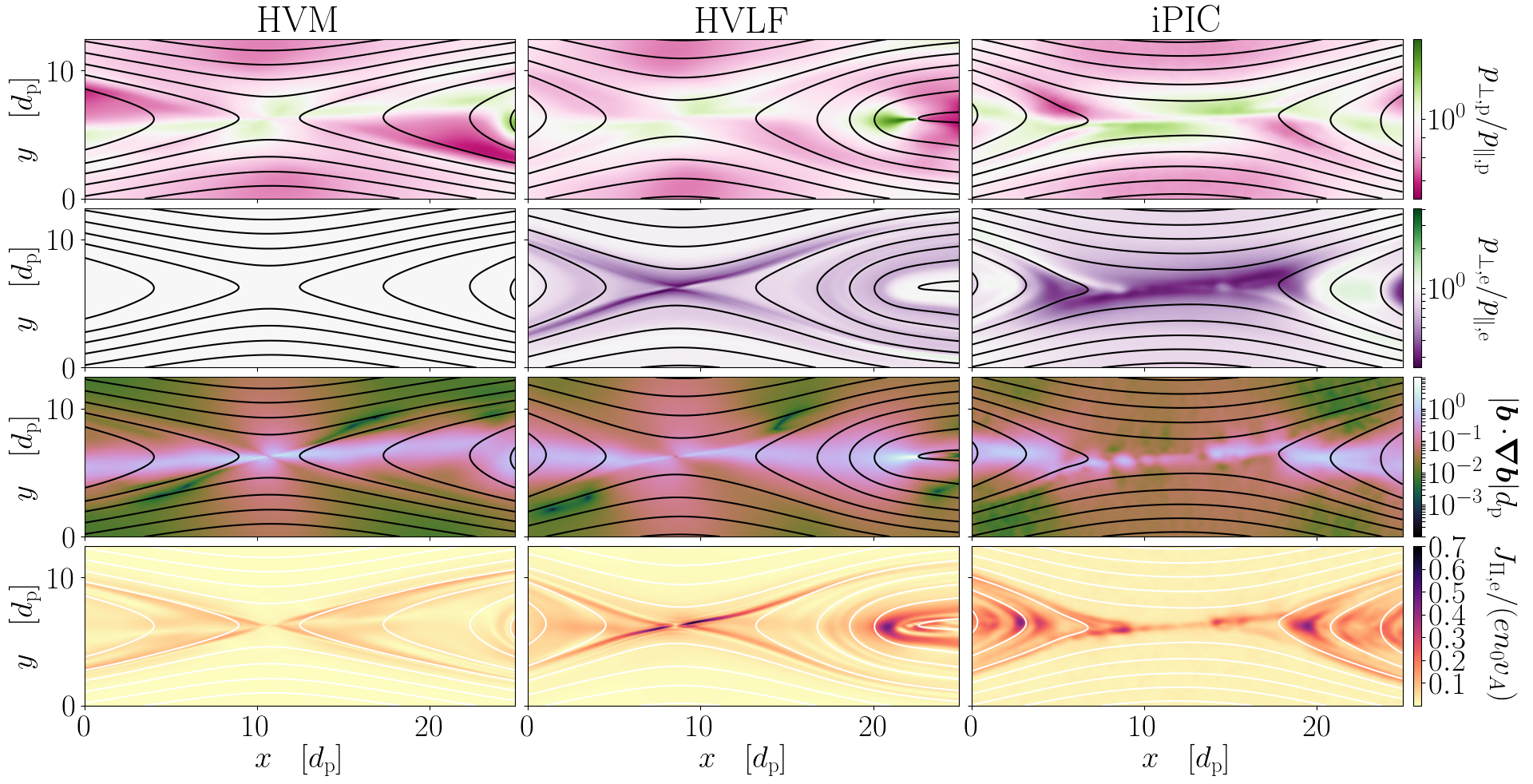}%
\caption{Heat maps of proton anisotropy (first row),  electron anisotropy (second row), magnetic curvature (third row), and $\bb{\nabla}\cdot\bb{\Pi}_{\mathrm{e}}$-drift current (fourth row) for the HVM (left column, $t=287.5\Omega_{\rm p}^{-1}$), HVLF (center column, $t=305.0\Omega_{\rm p}^{-1}$), and iPIC (right column, $t=320.0\Omega_{\rm p}^{-1}$) simulations. The three instants are chosen by fixing the value of the reconnected flux. This value is chosen in the quasi-stationary phase. The plot is zoomed-in around the main X point.}%
\label{fig:2d_anis_nonlin_zoom}%
\end{figure*}

\paragraph{\textbf{Electron-anisotropy effects on EDR shape.}} The morphology of the region around the X point is a consequence of the EDR elongation that arises both in the iPIC and HVLF simulations, but not in the HVM case. Although such an elongation is more pronounced in the iPIC simulation than it is for the HVLF one, this supports the idea that the elongation of the EDR is physical and associated with the local electron dynamics, the main reason being the electron pressure anisotropy. It is remarkable that a long time ago \citet{VasyliunasRvGSP1975} and, twenty years later, \citet{CaiPhPl1997}, pointed out that, during steady-state MR, the forces associated with the anisotropy in the electron pressure should play a significant role in balancing the electric field associated with reconnection in the neighborhood of the X point.
This idea was investigated more recently by \citet{LePhRvL2009,LeGeoRL2010,LePhPl2010} by means of a self-consistent model involving electron pressure anisotropies, Hall magnetic field, and the electrons jets. Inside and near the EDR, as shown in Fig.~\ref{fig:2d_nonlin_zoom}, a parallel electric field is generated as a consequence of the breakdown of the frozen-in condition.
This electric field enables efficient parallel electron heating and, together with a simultaneous adiabatic perpendicular cooling (due to gradients in the magnetic field), produces a significant amount of electron anisotropy, $p_{\parallel,\rm{e}}\gg p_{\perp,\rm{e}}$, typically observed in satellite data and confirmed in fully kinetic simulations. In our simulations, this anisotropy can be seen in the second row of Fig.~\ref{fig:2d_anis_nonlin_zoom}. Here, we observe a difference between the iPIC and the HVLF simulations. In particular, in the iPIC simulation the electron anisotropy affects a larger area. This is consistent with the idea of heating driven by electron trapping \citep[see][and references therein]{EgedalPhPl2013}. The kinetic trapping process is obviously completely lost in the electron isothermal HVM model but also in the extended HVLF description because of the fluid approach. On the other hand, it is remarkable that the anisotropy level reached in the HVLF simulation is comparable to that measured in the iPIC simulation. This is in agreement with the observed larger $E_\|$ structure in the HVLF and iPIC simulations. It is then the anisotropy of the electrons which in turn helps to balance the enhanced parallel electric field in the generalized Ohm's Law.

Effects of electron anisotropy can be seen also in the momentum balance. Neglecting electron inertia and assuming a gyrotropic $\bb{\Pi}_{\rm{e}}$, electrons momentum balance requires a perpendicular current given by
\begin{equation}\label{eq:J_perp_e}
    \bb{J}_{\perp,\mathrm{e}}\approx-ne\frac{\bb{E}\btimes\bb{B}}{B^2}+
    \frac{\bb{B}}{B^2}\btimes\grad p_{\perp,\mathrm{e}}+
    \frac{\bb{B}}{B^2}\btimes(p_{\parallel,\mathrm{e}}-p_{\perp,\mathrm{e}}){\bb{b}}\bcdot\grad{\bb{b}}\;.
\end{equation}
The first term on the right-hand side is the $\bb{E}\btimes\bb{B}$-drift. The two other drifts are driven by the pressure-gradient term $\bb{\nabla}\cdot\bb{\Pi}_{\rm{e}}$ (namely, the diamagnetic and the curvature drifts, respectively). In the region around the X point where electron jets develop, the last term of Eq.~\eqref{eq:J_perp_e} dominates over the $\bb{E}\btimes\bb{B}$-drift due to the high anisotropy and strong curvature. In Fig.~\ref{fig:2d_anis_nonlin_zoom} the curvature $|{\bb{b}}\cdot\bb{\nabla}{\bb{b}}|$ is shown in the third row, while the $\bb{\nabla}\cdot\bb{\Pi}_{\rm{e}}$-drift ($J_{\Pi,\mathrm{e}}$) is shown in the fourth one. $J_{\Pi,\mathrm{e}}$ turns out to be much larger in the HVLF and iPIC simulations as compared with the HVM one, and since the curvature does not show significant differences between the three runs, the difference can be only due to the electron anisotropy. As shown in the figure, this $J_{\Pi,\mathrm{e}}$ drift-current has clearly the greatest influence on $J_{\mathrm{e}}^{(\text{in-plane})}$ (cf. Fig.~\ref{fig:2d_nonlin_zoom}). 

This resulting anisotropy-dominated current is thus the main responsible for the asymmetric Hall magnetic field $\bb{B}_{\rm{H}}$ observed in HVLF and iPIC simulations. Another consequence is that the $\bb{J}\btimes\bb{B}$ force becomes dominant\footnote{Equation \eqref{eq:J_perp_e} implies $J_{\perp,\rm{e}}\sim n\,e\,v_{\rm{th,e}}\sim n\,e\,\sqrt{\beta_{\rm e}\,}\,v_{\rm A}$. When comparing the Hall force associated with such a current, $|\bb{J}\btimes\bb{B}|\sim n\,e\,\sqrt{\beta_{\rm e}\,}\,v_{\rm A}\,B$, with the force associated with the reconnecting electric field, $n\,e\,E_{\rm{rec}}\sim n\,e\,(0.1\,v_{\rm{A}}\,B)$, one finds that $|\bb{J}\btimes\bb{B}|\gg n\,e\,E_{\rm{rec}}$ for the typical values of $\beta_{\rm e}$ investigated in the present work.}. In order to understand what balances this force, the steady-state electron momentum equation can be cast in the form
\begin{equation}\label{eq:force_balance}
    \grad\bcdot\left[\left(B^2/2+p_{\perp,\mathrm{e}}\right)\bs{\mathrm{I}}+\left(p_{\parallel,\mathrm{e}}-p_{\perp,\mathrm{e}}-B^2\right){\bb{b}}{\bb{b}}\right]+\bb{F}=0\;,
\end{equation}
with $\bb{F}$ representing the other contributions (electric field, inertia, non-gyrotropy).
Where the magnetic-field curvature becomes important (also thanks to $\bb{B}_{\rm{H}}$), the term resulting from $\grad\bcdot({\bb{b}}{\bb{b}})$ can only be balanced by decreasing the quantity $|p_{\parallel,\mathrm{e}}-p_{\perp,\mathrm{e}}-B^2|$ (i.e., by increasing $p_{\parallel,\mathrm{e}}-p_{\perp,\mathrm{e}}$). Thus the electron pressure anisotropy acts as to roughly balance the magnetic tension, which is the equivalent of saying that $\bb{J}\btimes\bb{B}\approx\bb{\nabla}\cdot\bb{\Pi}_{\rm{e}}$. To summarize, anisotropy and B-curvature enhance electrons outflow jets, which generate the Hall magnetic field, $\bb{B}_{\rm{H}}$. This Hall component simultaneously provides the major contribution to the magnetic-field curvature, which in turn requires strong anisotropy to achieve force balance. This seems to explain why extended narrow electrons jets forms in models that include the electrons pressure anisotropy.

\subsection{Anisotropy}\label{subsec:anisotropy}

During the evolution of MR, we expect anisotropy to develop and to play a relevant role not only in the dynamics and morphology of the EDR around the X point (as outlined at the end of the previous subsection) but also by representing a source of free energy for possible secondary instabilities.

\begin{figure*}
    \centering%
    \includegraphics[width=1.0\textwidth]{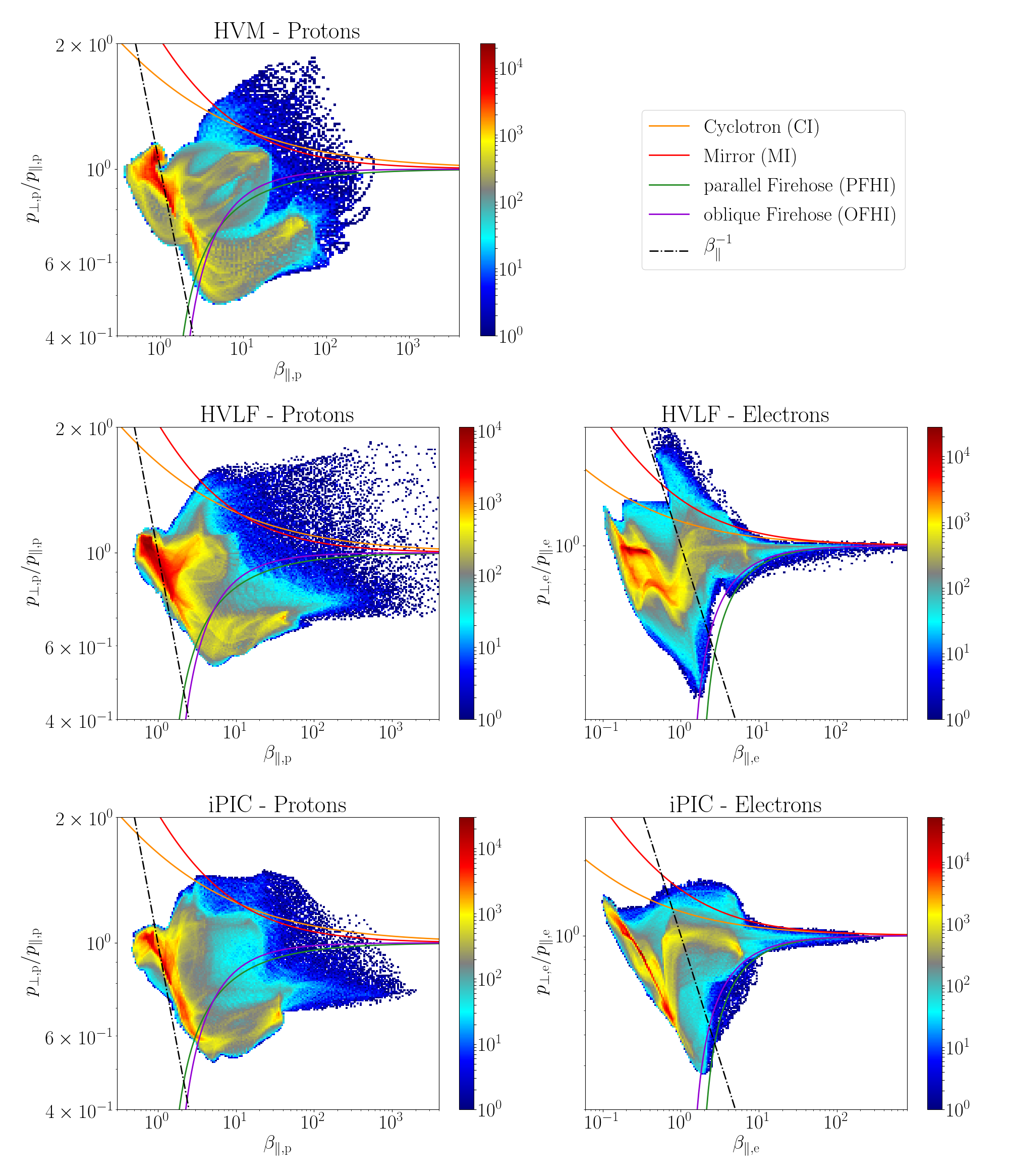}%
    \caption{Cumulative histogram of the $(\beta_{\parallel,\mathrm{a}}, T_{\perp,\mathrm{a}}/T_{\parallel,\mathrm{a}})$ occurrences over the quasi-stationary phase, for protons (left) and electrons (right) in the HVM (top, $t\in[262.5,\,287.5]\,\Omega_{\rm p}^{-1}$), HVLF (middle, $t\in[262.5,\,305]\,\Omega_{\rm p}^{-1}$), and iPIC (bottom, $t\in[275,\,320]\,\Omega_{\rm p}^{-1}$) simulations. The spatial region considered is defined by $y\leq12.5d_{\mathrm{p}}$. Solid curves represent different instability thresholds (the same color is adopted for both the proton and electron versions): cyclotron instabilities (orange), mirror instabilities (red), parallel fire-hose instabilities (green), and oblique fire-hose instabilities (violet). The black dash-dotted line is $1/\beta_{\parallel,\mathrm{a}}$.}
    \label{fig:windlike}
\end{figure*}

In Fig. \ref{fig:windlike} we show an histogram of the plasma distribution in a parameter space described by pressure anisotropy, $A_{\rm a}=p_{\perp,{\rm a}}/p_{\parallel,{\rm a}}$, versus parallel plasma beta, $\beta_{\parallel,\mathrm{a}}=8\pi p_{\parallel,\mathrm{a}}/B^2$, with ${\rm a}={\rm p}, {\rm e}$. 
Such a distribution is obtained by considering only the plasma within a region of the simulation that includes the active current sheet (roughly one third of the simulation box, in height). Also, we take the cumulative data over a time interval in the quasi-stationary phase\footnote{Since we `synchronize'' the simulations via the reconnected magnetic flux, these intervals differ from run to run and are reported in the description of Fig.~\ref{fig:windlike}.}. This is a typical representation of the plasma distribution used, for instance, by SW turbulence studies~\citep[e.g.,][]{HellingerGeoRL2006,MatteiniGRL2007,BalePRL2009}.
In these plots, we also report a set of curves representing the thresholds for anisotropy-driven instabilities, listed in the top right box. The upper branches ($p_{\perp,\mathrm{a}} > p_{\parallel,\mathrm{a}}$ region) are the MI (red) and the cyclotron instability (CI; orange); lower branches are parallel FHI (PFHI; green) and oblique FHI (OFHI; violet). The same color is used for both the protons and the electrons versions. These curves, used for reference, are taken from different works. For what concerns protons, the CI threshold is from \citet{LazarMNRAS2014}, while the MI and FHI curves are taken from \citet{MarucaApJ2012} and \citet{AstfalkJGRA2016}, respectively. The electron version for these thresholds are instead taken from \citet{LazarA&A2013} (CI), \citet{GaryJGRA2006} (MI), \citet{GaryPhPl2003} (PFHI), and \citet{HellingerJGRA2014} (OFHI). 
In order to be consistent with the duration of our simulations, all the proton-instability thresholds, as well as the electron PFHI curve, are computed for the maximum growth rate $\gamma_m/\Omega_{\rm p}^{-1}=10^{-2}$. The remaining electron-instability thresholds are computed for $\gamma_m/\Omega_{c,\mathrm{e}}^{-1}=10^{-3}$ (which, in our simulations, means $\gamma_m/\Omega_{\rm p}^{-1}=10^{-1}$).

\paragraph{\textbf{Proton anisotropy.}} By comparing the distribution of protons in HVM, HVLF, and iPIC simulations in Fig.~\ref{fig:windlike}, one can see that they do not behave qualitatively very differently. That is true also in the spatial domain, as presented in Fig.~\ref{fig:2d_anis_nonlin_zoom}.  In all the three cases (in Fig.~\ref{fig:windlike}), a significant fraction of protons are indeed exceeding the FHI thresholds. However, a difference in proton anisotropy can be seen in the HVM simulation, where there is a larger fraction of protons that reaches values of pressure anisotropy that are further below $1$, with respect to the HVLF and iPIC cases. In fact, when electrons are isothermal, the free energy available in the system can only feed the protons' anisotropy. On the other hand, when electron anisotropy is allowed, such free energy is shared between the two species. This suggests that the electron model implemented in the HVLF simulation is capable of distributing free energy among the two species in a way similar to the full-kinetic case.

\begin{figure*}
    \centering%
    \includegraphics[width=1.0\textwidth]{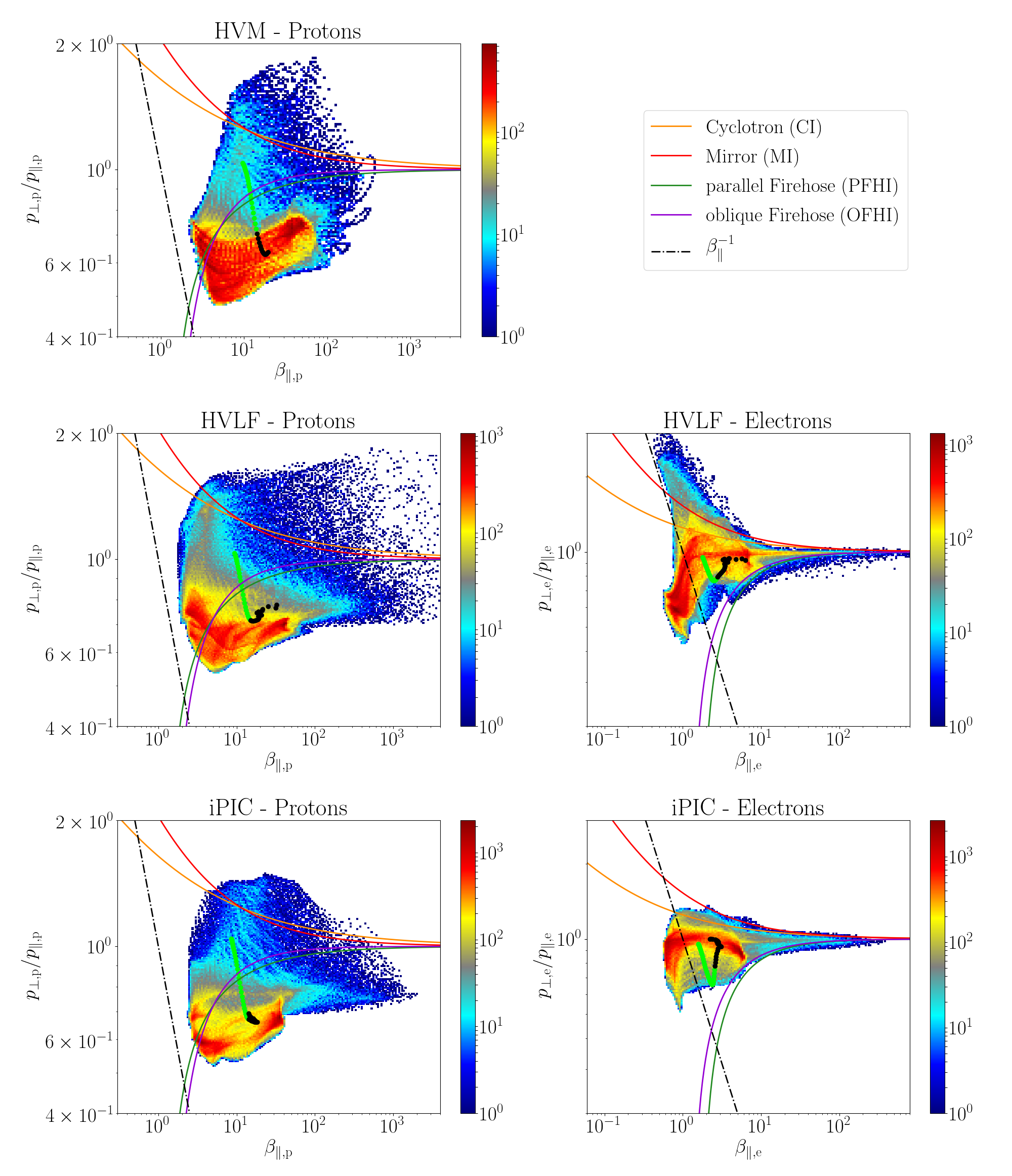}%
    \caption{Cumulative histogram of the $(\beta_{\parallel,\mathrm{a}}, T_{\perp,\mathrm{a}}/T_{\parallel,\mathrm{a}})$ occurrences over the quasi-stationary phase for protons (left) and electrons (right) in the HVM (top, $t\in[262.5,\,287.5]\,\Omega_{\rm p}^{-1}$), HVLF (middle, $t\in[262.5,\,305]\,\Omega_{\rm p}^{-1}$), and iPIC (bottom, $t\in[275,\,320]\,\Omega_{\rm p}^{-1}$) simulations. The spatial region considered is the island interior, here defined by $\Psi/\max(\Psi) > 0.9$. The full circles on top of the histogram represent the time evolution of the island-averaged anisotropy and parallel beta: black circles denote times over which the cumulative histogram is computed, while green circles represent earlier times. Curves are the same as in Fig.~\ref{fig:windlike}.}
    \label{fig:island_in_wind}
\end{figure*}
%
The main question remains why a non-negligible fraction of protons populate a region beyond the FHI marginal stability.
In order to elucidate this point, we consider the spatial region confined within a magnetic island: analogously to Fig.~\ref{fig:windlike}, in Fig.~\ref{fig:island_in_wind} we show a time-cumulative histogram where now only the plasma inside the magnetic island is considered (islands elements are selected using a threshold on the magnetic flux, namely $\Psi / \max(\Psi)> 0.9$). The full circles reported on top of the plasma distribution represent a time series of the instantaneous values of anisotropy and parallel beta, averaged over the island domain, {\em viz.} $(\langle p_{\perp,\mathrm{a}}/p_{\parallel,\mathrm{a}}\rangle_{island},\langle\beta_{\parallel,\mathrm{a}}\rangle_{island})$ versus time. The black circles denote the time range over which the cumulative histogram is computed, while green circles represent earlier simulation times.
From Fig.~\ref{fig:island_in_wind}, one clearly appreciates that the interior of the magnetic island (or, at least, part of it) is where the proton population that fills the FHI-unstable regions comes from. 
On top of that, one clearly appreciates that there is a noticeably greater portion of protons that is beyond the FHI thresholds in the HVM simulation if compared with the HVLF and iPIC cases. In fact, there is a larger fraction of the proton distribution, especially in its low-$\beta_{\parallel,\mathrm{p}}$ portion, that is found in the unstable region for the HVM case. This is highlighted also by the island-averaged points, which, in the HVM simulation, reach the lowest values among the three models. These features of the proton anisotropy-beta distribution within the various models is (one of) the effect of how adopting a different electron response feeds back onto the proton dynamics (namely, whether or not electrons can develop pressure anisotropy, and thus distribute the initial free energy among more available energy channels; see, e.g., \citealt{CerriCamporealePOP2020}).

Complementary to Fig.~\ref{fig:island_in_wind}, we also performed an analysis of the spatial locations of the plasma elements out of thresholds within the simulation domain, using the whole distribution reported in Fig. \ref{fig:windlike} (not shown here). We find confirmation that the islands are the main origin of the proton population that is found in the parameter region unstable to the OFHI, although part of the outflows also contributes. At the same time, there are spatial regions that become unstable for the proton CI and which are similar in all three simulations: These regions are correlated with points where $B_y$ accumulates due to the reconnection process.
Finally, adjacent to these latter spatial domains, there are regions where protons are MI-unstable.

The fact that the island is the unstable region is not surprising since two factors are simultaneously at play: (i) islands are located in a low magnetic field region, which favors the instability development (i.e., decreasing $T_\perp$ due to magnetic-moment conservation), and (ii) outflows and compression enhance parallel heating (i.e., increasing $T_\|$). The question is rather why this unstable region is not regulated by the instability (i.e., releasing anisotropy-related free-energy as, for instance, Alfv\'en waves, and bringing the distribution back to marginal stability). The FHI thresholds shown in the figures are computed for a (maximum) growth rate of $10^{-2}\Omega_{\rm p}$, so that in about $100\Omega_{\rm p}^{-1}$ one should see, at least, a reduction of this unstable region. Our explanation is that the limited dimension of the islands (length $<40d_{\mathrm{p}}$) prevents the formation of modes at $k\,\rho_{\mathrm{p}} < 1$, corresponding to the maximum growth rate for the FHI \citep{HellingerJGR2000,SchekochihinMNRAS2010}. For $k_{min}=2\pi/\lambda_{max}\approx 2\cdot10^{-1}d_{\mathrm{p}}^{-1}$ and $\rho_{\mathrm{p}} = \sqrt{\beta_{\mathrm{p}}/2}\, d_{\mathrm{p}}\approx 5 d_{\mathrm{p}}$ (see Fig.~\ref{fig:island_in_wind}), one finds that $k_{min}\,\rho_\mathrm{p} \approx 1$. We conclude that the fastest-growing modes associated with these FHI branches cannot emerge due to the spatial constraints imposed by the islands size.

In Fig.~\ref{fig:island_in_wind}, dash-dotted black lines are $\propto 1/\beta_{\parallel,\mathrm{a}}$ curves, presented here to show how both electrons and protons populating the island seem to follow, on average, a double-adiabatic evolution \citep[i.e., close to a CGL model with zero heat fluxes; see][]{MatteiniSSRv2012,HunanaJPlPh2019a}, until $t\approx 270 \Omega_{\rm p}^{-1}$.
Of course, since proton FHI is geometrically suppressed, nothing confines them during this expansion, which continues until the quasi-stationary phase (i.e., the black circles). Nevertheless, in all three simulations, the protons average point seems unable to go back to marginal stability at later times.

\paragraph{\textbf{Electron anisotropy.}} We now discuss the effectiveness of the HVLF model in capturing (as possible) the electrons dynamics in the $(p_{\perp,\rm{e}}/p_{\parallel,\rm{e}},\beta_{\parallel,\rm{e}})$ plane, when compared to the full-kinetic case. In both HVLF and iPIC simulations, the electron distribution at $p_{\perp}/p_{\parallel}<1$ is indeed well bounded by the FHI marginal stability thresholds (see Fig.~\ref{fig:windlike}). In particular, one can see that the distribution lower boundaries are slightly better in agreement with the OFHI than with the PFHI (besides, we remind the reader that the threshold for the parallel branch is computed for a maximum growth rate that is ten times smaller than its oblique counterpart). In Fig.~\ref{fig:island_in_wind}, on the side of the electrons (right column), we can appreciate the confinement action of the electron OFHI within the island. Indeed, during its time evolution, the island-averaged point roughly follows a double-adiabatic trajectory, approaching the electrons FHI thresholds, and then it bounces back within the stable region. After that, from both the average points and the histograms, we see a tendency toward the electrons MI marginal stability threshold. A full (nonlinear) development of mirror modes should be associated with the formation of magnetic holes~\citep[or magnetic peaks, see][and references therein]{HellingerJPlPh2018}. However, in our simulations, the presence of the MR process (and the associated magnetic structures) does not allow us to clearly identify the above-mentioned mirror-related magnetic structures.

A similar degree of confinement is observed at high $p_{\perp,\rm{e}}/p_{\parallel,\rm{e}}$ and high $\beta_{\parallel,\rm{e}}$ values, where the system is well restrained by the electrons MI for $\beta_{\parallel,\rm{e}}>1$. On the other hand, at lower $\beta_{\parallel,\rm{e}}$ values, where the electrons CI should be responsible for the confinement, the HVLF models (obviously) fails. This confinement is instead present in the fully kinetic model. Indeed, even if in Fig.~\ref{fig:windlike} a small portion of the electrons are unstable also in the iPIC run, we clarify that this is due to a transient (namely, a plasmoid) observed in this simulation. This structure lasted for about $30\Omega_{\rm p}^{-1}$, with a dynamics ten times faster than the one associated with the X-point reconnection. At this timescale (given the mass ratio implemented in our simulations), even the electrons ``lag behind'' the produced anisotropy. This anisotropy is quickly regulated after the plasmoid is ``absorbed'' within the main island. As a check for a link between this anisotropy and the plasmoid, we can show that it is absent in the island, as seen in Fig.~\ref{fig:island_in_wind}. We also checked that the portion of electrons that are unstable to CI in the iPIC case is, indeed, spatially located within the plasmoid.

This result can be considered as a physiological difference between the electron models implemented in the HVLF and iPIC simulations (i.e., due to the lack of electron-FLR effects in the HVLF model). On top of that, we note that the portion of electron distribution populating the CI-unstable region in the HVLF simulation follows the $1/\beta_{\parallel,\mathrm{e}}$ double-adiabatic evolution \citep[see][]{MatteiniSSRv2012,HunanaJPlPh2019a}. Moreover, we checked that this unstable portion of the electrons in the HVLF case is, again, inside the island. This can be viewed as one more signature that processes occurring within a magnetic island, such as island contraction, contribute significantly to the anisotropy generation in both directions (i.e., not only pushing the plasma toward FHI regimes). Overall, this corroborates the hypothesis that the differences in electrons CI confinement between HVLF and iPIC simulations are physical rather than numerical. We also mention that indeed the iPIC simulation presents a small amount of electron cooling due to the implicit algorithm: so, one may think that this cooling partially dissipates the free-energy that would later lead to high-$p_{\perp,\rm{e}}/p_{\parallel,\rm{e}}$ anisotropies. Being aware of this numerical effect, we checked the behavior of internal, kinetic, and total energies (i) of the single species and (ii) of the whole plasma, during the iPIC simulation (not shown). Even if a certain amount of numerical cooling is occurring (mostly noticeable at early times, and at the expense of electrons internal energy), that amount is not enough to explain the differences just discussed between the iPIC and the HVLF runs, especially in the quasi-steady stage. Therefore, we confirm that the lack of plasma population above the electrons CI threshold in the iPIC simulation is the result of a physical, rather than numerical, effect.

\begin{figure}[!ht]
    \centering%
    \includegraphics[width=.5\textwidth]{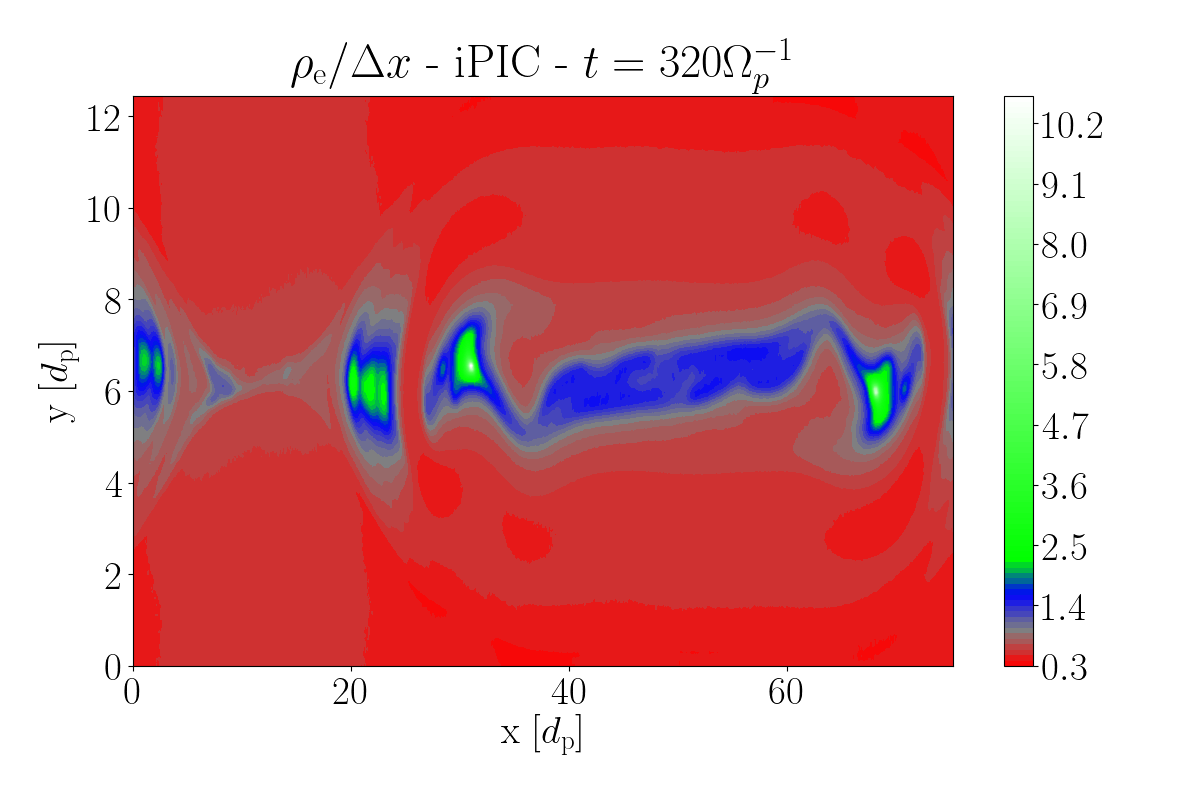}%
    \caption{Heat map of the local electron Larmor radius (normalized to the grid spacing) from the iPIC run, at $t=320\Omega_{\rm{p}}^{-1}$. The color map is suited for a qualitative analysis: gray for $\rho_e/\Delta x\approx1$, red for $\rho_e/\Delta x < 1$, blue for $1 < \rho_e/\Delta x < 2$, and green for $\rho_e/\Delta x > 2$.}
    \label{fig:rho_e}
\end{figure}

The iPIC and the HVLF models, as expected, support in some cases different dynamics, in particular for low $\beta_{\|,\rm{e}}$ ($\sim1$) and high electron anisotropy ($>1$). This discrepancy is due to the possibility of the iPIC model capturing electron anisotropy-limiting processes that the HVLF cannot. Such kinds of processes can be reproduced only if the electron relevant scales (i.e., the electron Larmor radius in space and the electron gyration in time) are resolved in the model. In our iPIC simulation, we duly resolve electron gyration in time but not in space at the initial time (as discussed in Sect.~\ref{subsec:setup}).
However, due to the system evolution, the electron gyroradius increases locally, becoming larger than the grid spacing in a few extended regions.
Figure~\ref{fig:rho_e} shows a 2D map of the ratio between $\rho_{\rm{e}}$ and $\Delta x$, in which green and blue denote regions where this ratio is grater than one. Those regions present a higher perpendicular temperature that corresponds to a larger value of the local Larmor radius. It is in those regions that electron FLR effects are captured by the iPIC model. The consequence of these effects is to limit the further development of $p_{\perp,\rm{e}}>p_{\|,\rm{e}}$ anisotropy, which, to the contrary, is not limited by these effects in the HVLF case.
In conclusion, electron gyration scale effects are at play in the iPIC run and not in the HVLF one, resulting in different electron anisotropy dynamics in the two cases.

\begin{figure*}
    \centering%
    \includegraphics[width=.99\textwidth]{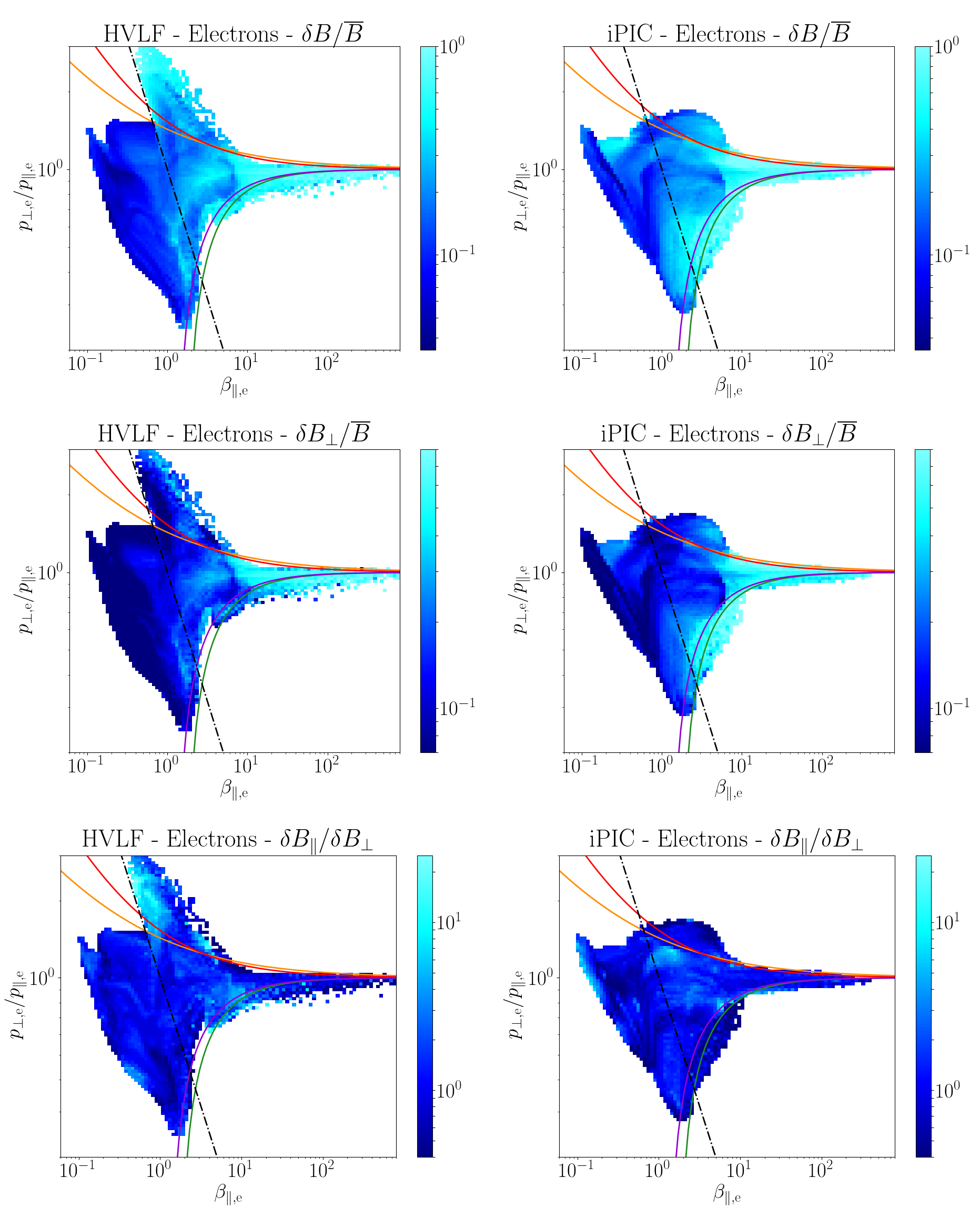}%
    \caption{Cumulative, bin-averaged histogram of $\delta B/\overline{B}$ (top row), $\delta B_{\perp}/\overline{B}$ (middle row), and $\delta B_{\parallel}/\delta B_{\perp}$ (bottom row) in the $(\beta_{\parallel,\mathrm{e}}, T_{\perp,\mathrm{e}}/T_{\parallel,\mathrm{e}})$ plane. Histograms are cumulated over the quasi-stationary phase ($t\in[262.5,\,305]\,\Omega_{\rm p}^{-1}$ for HVLF, left column; $t\in[275,\,320]\,\Omega_{\rm p}^{-1}$ for iPIC, right column). Here, $\delta B = |\delta\bb{B}|\equiv|\bb{B}-\langle\bb{B}\rangle_t|$ and $\overline{B}=|\langle\bb{B}\rangle_t|$, where $\langle \dots\rangle_t$ denotes the time average over the interval used for the histogram. The spatial region considered and the curves are the same as in Fig.~\ref{fig:windlike}.}
    \label{fig:fluct_in_wind}
\end{figure*}

\paragraph{\textbf{Character of magnetic fluctuations.}} In order to further investigate the role of these instabilities in the regulation of the electrons' anisotropy distribution, we also analyze the properties of magnetic fluctuations in the quasi-stationary phase, in a similar way to what is done in \citet{BalePRL2009} for SW turbulent fluctuations. 
Such an analysis is reported in Fig.~\ref{fig:fluct_in_wind}, where the properties of magnetic-field fluctuations are reported in a $(\beta_{\parallel,\mathrm{e}}, p_{\perp,\mathrm{e}}/p_{\parallel,\mathrm{e}})$ plane: this is done relating the fluctuations' values at any spatial point to the anisotropy and parallel beta values of the plasma at that same position. 
(The curves and time intervals considered here are the same for Fig.~\ref{fig:windlike}, although a slightly more coarse-grained binning is employed for these plots.) 
The color-map refers to the bin-averaged value of $\delta B/\overline{B}$ (top row), $\delta B_{\perp}/\overline{B}$ (middle row), or $\delta B_{\parallel}/\delta B_{\perp}$ (bottom row), measured in the HVLF (left column) and iPIC (right column) simulations. 
Here, $\delta B = |\delta\bb{B}|\equiv|\bb{B}-\langle\bb{B}\rangle_t|$ and $\overline{B}=|\langle\bb{B}\rangle_t|$, where $\langle \dots\rangle_t$ represents time average over the range that is being considered. 
If just the magnitude of magnetic-field fluctuations are considered (i.e., $\delta B/\overline{B}$; Fig.~\ref{fig:fluct_in_wind}, top row), one finds that the larger fluctuations are mainly located in the regions nearby marginal-stability thresholds, for both $p_{\perp,\mathrm{e}}/p_{\parallel,\mathrm{e}}>1$ and $p_{\perp,\mathrm{e}}/p_{\parallel,\mathrm{e}}<1$ values of anisotropy. 
Such a fluctuation enhancement is interpreted as a signature of the growth of these anisotropy-driven instabilities in their attempt to regulate electron-pressure anisotropy: larger fluctuations tend to scatter electrons more efficiently, so to isotropize their pressure and bring the plasma back to marginal stability. 
Considering now the magnetic fluctuations transverse to the mean field, $\delta B_{\perp}/\overline{B}$ (Fig.~\ref{fig:fluct_in_wind}, middle row), one finds that these type of fluctuations are strongly enhanced across the FHI marginal stability (at $p_{\perp,\mathrm{e}}/p_{\parallel,\mathrm{e}}<1$), and nearby the MI and CI thresholds (at $p_{\perp,\mathrm{e}}/p_{\parallel,\mathrm{e}}<1$, but only for those regions where $\beta_{\parallel,\mathrm{e}}\gtrsim10$). 
These features are slightly clearer in the iPIC simulation than they are in the HVLF case. 
Indeed, in the HVLF simulation, the electrons do not clearly reach and/or overcome the FHI thresholds at $(\beta_{\parallel,\mathrm{e}}<10$. 
As mentioned above, this difference is due to the presence of a transient plasmoid structure in the iPIC simulation (cf. Figs.~\ref{fig:windlike} and \ref{fig:island_in_wind}).

Considering now the so-called magnetic compressibility of the fluctuations, $\delta B_{\parallel}/\delta B_{\perp}$ (Fig.~\ref{fig:fluct_in_wind}, bottom row), both simulations exhibit an enhancements of this fluctuations' property near the MI (and/or CI) thresholds, when  $1\lesssim\beta_{\parallel,\mathrm{e}}\lesssim10$.
Magnetic-fluctuations compressibility is also found to be enhanced near the FHI thresholds, and in particular around the oblique branch, for those regions where $5\lesssim\beta_{\parallel,\mathrm{e}}\lesssim100$.
What is often called OFHI should be better called non-propagating FHI, since at oblique propagation the fire-hose splits into a propagating, $\omega_{\rm{r}}\neq0$, and a non-propagating, $\omega_{\rm{r}}=0$, branches (here, $\omega_{\rm r}$ is the real-part of the complex frequency of the mode, $\omega=\omega_{\rm r}+i\gamma$). The $\omega_{\rm{r}}\neq0$ branch is a continuation (at nonzero propagation angles) of the parallel case, it is a transverse (non-compressional) mode, and it resonates with protons, not with electrons. The non-propagating branch instead is (at least partially) compressive and dominates over the propagating one. Focusing on the $\omega_{\rm{r}}=0$ branch, when the compressive component is small (i.e., $\delta B_{\parallel}/\delta B_{\perp}<1$), the instability is cyclotron resonant with electrons, and for this reason is able to control the electron anisotropy. When $\delta B_{\parallel}/\delta B_{\perp}>1$, this mode becomes an analogous to the electron-mirror mode, just on the other ``side'' of the anisotropy spectrum \citep[see][and references therein]{LazarSoPh2014}.

Of course, it is important to stress also the limits of the HVLF model. We already mentioned the inability of the LF closure to reproduce the electrons CI. In the bottom row of Fig.~\ref{fig:fluct_in_wind}, from the iPIC simulation we observe a very small magnetic compressibility for $\beta_{\parallel,\mathrm{e}}<1$, especially near the instability thresholds at $p_{\perp,\mathrm{e}}/p_{\parallel,\mathrm{e}}>1$. This is likely due to the fact that electrons anisotropy is efficiently confined by their CI in the iPIC simulation. On the other hand, it is clear that the HVLF model cannot account for the electron CI process: the high magnetic-compressibility values in this region thus suggest that is the MI that tries to make up for the absence of the CI in regulating the electron-pressure anisotropy (with little success).

\section{Discussion and conclusions}\label{sec:discussion}

In this study we performed 2D numerical simulations of MR within a typical Harris's sheet configuration.
The same initial configuration is evolved by means of three different plasma models: (i) an HVM model with (isotropic) isothermal electrons with mass, (ii) an HVLF model, where an anisotropic electron fluid is equipped with an LF closure, and (iii) a full-kinetic model employing an iPIC algorithm.
The goal of this work is to investigate to which extent different reduced electron models are able to capture the physics of MR, with the focus on its nonlinear stage. In particular, we are interested in elucidating the advances represented by the HVLF model, with respect to a standard HVM approach, in the capability of capturing several aspects of the electron dynamics and its feedback onto the protons.
These aspects include the development of electron-pressure anisotropy and anisotropy-driven (electron) instabilities, as well as the electron-kinetic response, such as Landau damping.
Limitations that still hold within an HVLF approach with respect to a full-kinetic model are also highlighted.

As far as the global reconnection dynamics, expressed by the reconnected flux and its reconnection rate,  is concerned, all three simulations show results that are qualitatively in agreement with one another.
In particular, after an initial linear stage exhibiting minor differences between the HVM model and the other two models, the three simulations achieve a very good agreement in the so-called super-exponential phase. 
In the quasi-stationary stage, on the other hand, there is a quantitative difference in the quasi-steady reconnection rate between HVM and the other two models that is worth mentioning: The reconnection rate in the quasi-steady regime is noticeably higher in the HVM model, about $\approx0.06$, than in the HVLF and iPIC simulations, where it sets around $\approx0.04$. 
This reconnection rate value  differs from the one considered to be a ``typical'' value for fast collisionless reconnection, $\approx0.1$. Such a difference in the quasi-steady reconnection rate is due to a background-to-peak density ratio that in our configuration is higher than the one that is ``typically'' implemented~\citep[see, e.g.,][and references therein]{CassakJPlPh2017}.
As for the local features of the dynamics, the absence of electron anisotropy in the HVM simulation results in a different shape of the region around the X point of the reconnecting current sheet and, in particular, of the EDR. 
From the X point, in a region surrounding the EDR, elongated and ``tilted'' electron jets develop as a consequence of the electron-pressure anisotropy that affects the momentum balance in the HVLF and iPIC simulations. 
For the same reason, the current structure and the EDR are also elongated (and the former is also tilted), a feature that is not observed with isotropic isothermal electrons.

Then, we focused on the production and regulation of species' pressure anisotropy during the nonlinear stage of reconnection. The processes mainly involved in the confinement of the plasma distribution in a anisotropy-versus-beta parameter space are the OFHI (for $p_{\perp,\mathrm{a}}<p_{\parallel,\mathrm{a}}$), the MI (for $p_{\perp,\mathrm{a}}>p_{\parallel,\mathrm{a}}$ and $\beta_{\parallel,\mathrm{a}}\gtrsim2$), and the CI (for $p_{\perp,\mathrm{a}}>p_{\parallel,\mathrm{a}}$ and $\beta_{\parallel,\mathrm{a}}\lesssim2$). 
As far as the dynamics of the protons is concerned, all three simulations are qualitatively similar. 
A significant portion of protons that are inside a magnetic island continuously populate the parameter-space region that is beyond the OFHI marginal-stability threshold. 
This is due to the size of the island not being long enough for the fastest growing mode of the instability to develop and to efficiently regulate their anisotropy (while island contraction simultaneously keeps generating more pressure anisotropy). 
Furthermore, when compared to the HVLF and iPIC simulations, protons in the HVM simulation show a tendency to develop larger values of pressure anisotropy (toward $p_{\perp,\mathrm{i}}<p_{\parallel,\mathrm{i}}$).
This is the consequence in the HVM model of adopting an isotropic pressure for the electrons, which excludes \emph{a priori} the possibility of feeding electron anisotropy using part of the energy released by the reconnection process.
However, we defer a detailed study of this aspect to future works.

As far as electron dynamics is concerned, we have shown that the HVLF and iPIC simulations lead to very similar results, although some differences between the two are still present. 
A key finding is that, when compared to the full-kinetic case, the anisotropic electron-fluid model with an LF closure efficiently regulates the distribution of electrons in an anisotropy-versus-beta plane. 
In fact, the OFHI and the MI are effective at controlling the electron anisotropy in both simulations. 
The main difference between HVLF and iPIC arises due to the intrinsic limitation of the electron-LF model, which cannot capture the electron CI. 
The CI threshold is indeed effective at limiting the electron-pressure anisotropy (at $p_{\perp,\mathrm{e}}>p_{\parallel,\mathrm{e}}$ and $\beta_{\parallel,\mathrm{e}}\lesssim2$) in the iPIC simulation, while in the HVLF case such anisotropy can only be regulated by the MI (which in our setting is less effective than CI, and thus a larger electron anisotropy is achieved). 
This means that, when electron-gyration effects are taken into account, such large anisotropy is not observed.
In this context, electron FLR effects could also play a relevant role in limiting the anisotropy.

In conclusion, we have employed three different models to describe the same process, focusing on the effectiveness of how the electron physics can be modeled within a hybrid-kinetic framework (namely, including electron-pressure anisotropy and a fluid model of their linear kinetic response). 
When compared to the full-kinetic case, the electron response implemented in the HVLF model effectively reproduces the main features of MR in 2D, as well as several aspects of the associated electron microphysics and its feedback onto protons dynamics. 
This includes the global evolution of MR and the local physics occurring within the EDR, as well as the evolution of species' pressure anisotropy. 
In particular, OFHI, MI, and CI play a relevant role in regulating electrons' anisotropy during the nonlinear stage of MR. 
As expected, the HVLF model captures all these features, except for the electron CI.
The HVLF model represents a good approximation of the full-kinetic case that compromises between the richness of the physics that is described and its computational cost.

Future works will extend this study to MR in 3D, as well as to plasma turbulence across the so-called transition range between MHD and sub-ion scales~\citep[where reconnection plays a crucial role; see, e.g.,][and references therein]{CerriCalifanoNJP2017}.
Regarding additional model development, one can further refine the electron-fluid model by including (large-scale) electron FLR effects and/or moving the LF closure to higher moments~\citep[see, e.g.,][]{SulemPassotJPP2015}. 
This would make the electron model even closer to a full-kinetic response, although at a somewhat higher computational cost, and is currently considered to be a possible future direction.

\begin{acknowledgements}
This project has received funding from the European
Union’s Horizon 2020 research and innovation program under grant agreement No. 776262 (AIDA, www.aida-space.eu).
F.F., S.S.C., and F.C. want to thank D.~Telloni for useful discussions regarding {\em in-situ} observations associated with magnetic reconnection.  
We acknowledge the Italian ISCRA initiative (grant HP10B10ALD), the European PRACE initiative (grant n.~2017174107 and n.~pn56ye) for awarding us access to the supercomputer Marconi, CINECA, Italy and to SuperMUC-NG at GCS@LRZ, Germany. We thank Dr. M. Guarrasi (CINECA, Italy) for his essential contribution to code implementation on Marconi.
S.S.C. is supported by the Max-Planck/Princeton Center for Plasma Physics (NSF grant PHY-1804048).
F.P. is supported by the PostDoctoral Fellowship 12X0319N and the Research Grant 1507820N from Research Foundation -- Flanders (FWO).
This work has been performed under the Project HPC-EUROPA3 (INFRAIA-2016-1-730897), with the support of the EC Research Innovation Action under the H2020 Programme; in particular, F. P. acknowledges the support of F. C. and The Department of Physics at the University of Pisa and the computer resources and technical support provided by CINECA.
\end{acknowledgements}

\bibliographystyle{aa}
\bibliography{biblio}

\end{document}